\documentclass[aps,prd,nofootinbib,reprint,twocolumn,longbibliography,superscriptaddress,eqsecnum,floatfix,scrartcl]{revtex4-1}
\pdfoutput=1      
\usepackage{lipsum}

\usepackage{amssymb}  
\usepackage{color,soul} 
\usepackage{graphicx}  
\usepackage[tight]{subfigure}  
\usepackage{mathrsfs}
\usepackage[pdftex,pdfusetitle,bookmarks=true,colorlinks=true,citecolor=blue,urlcolor=blue,linkcolor=magenta]{hyperref}
\usepackage{hypcap}
\usepackage{array}
\usepackage{multirow}
\usepackage{amsmath,amssymb}

\graphicspath{{./}{./images/}}

\begin{document}
\title{Quantum butterfly effect in polarized Floquet systems}    
\author{Xiao Chen}   
\affiliation{Department of Physics and Center for Theory of Quantum Matter, University of Colorado, Boulder, CO 80309, USA}
\affiliation{Department of Physics, Boston College, Chestnut Hill, Massachusetts 02467, USA}
\author{Rahul M. Nandkishore}   
\affiliation{Department of Physics and Center for Theory of Quantum Matter, University of Colorado, Boulder, CO 80309, USA}
\author{Andrew Lucas}   
\affiliation{Department of Physics and Center for Theory of Quantum Matter, University of Colorado, Boulder, CO 80309, USA}
\email{andrew.j.lucas@colorado.edu}

\date{\today} 

\begin{abstract}      
We explore quantum dynamics in Floquet many-body systems with local conservation laws in one spatial dimension, focusing on sectors of the Hilbert space which are highly polarized.   We numerically compare the predicted charge diffusion constants and quantum butterfly velocity of operator growth between models of chaotic Floquet dynamics (with discrete spacetime translation invariance) and random unitary circuits which vary both in space and time.   We find that for small but  nonzero density of charge (in the thermodynamic limit), the random unitary circuit correctly predicts the scaling of the butterfly velocity but incorrectly predicts the scaling of the diffusion constant.  We argue that this is a consequence of quantum coherence on short time scales.  Our work clarifies the settings in which random unitary circuits provide correct physical predictions for non-random chaotic systems, and sheds light into the origin of the slow down of the butterfly effect in highly polarized systems or at low temperature.
%
\end{abstract}

\maketitle   
\section{Introduction}
Understanding  
the approach to equilibrium
in strongly interacting many-body systems has become a problem of significant interest in the past few years.   It is widely expected that certain features of dynamics will be universal among many different quantum systems --  for example, the emergence of diffusion and hydrodynamics in systems with conserved quantities \cite{kadanoffmartin},  or the quantum butterfly effect in which the domain of support of operators expands ballistically in systems with sufficiently local interactions \cite{Lieb1972}. Much recent work has accordingly focused on cartoon models of quantum dynamics consisting of random unitary gates applied in discrete time steps:  the random unitary circuit (RUC) \cite{Nahum2017, nahum_operator_2018,von_keyserlingk_operator_2018, rakovszky_diffusive_2017, khemani_operator_2017}. 
These RUCs often provide an analytical solution for many-body quantum dynamics and, it is hoped, shed light on the dynamics for more general quantum systems.  Yet in all these RUC models, a key feature is randomness in both time and spatial directions. Averaging over this randomness leads to substantial decoherence in the quantum unitary evolution and, in many simple cases, maps quantum dynamics to a classical stochastic process.  However, in systems without randomness, it is less clear whether quantum coherence is as negligible as RUCs suggest, and it is important to learn whether and when non-random chaotic systems have different dynamical behavior than RUCs. 
Indeed, there are simple models where such random circuits fail to properly describe chaotic Hamiltonian quantum evolution \cite{Lucas2019_star}.

In this paper, we focus on a specific question:  do random circuits correctly describe the growth of operators in one dimensional Floquet systems?  The operator growth can be (partially) diagnosed by the out-of-time-ordered correlator (OTOC)\cite{larkin_quasiclassical_1969,maldacena_bound_2015},   
\begin{align}
C(r,t)=-\mbox{Tr}\left\{[O_1(0,t), O_2(r,0)]^2 \right\}/\mbox{Tr}(\mathbb{I}).  \label{eq:otocintro}
\end{align}
where $\mbox{Tr}(\mathbb{I})$ is the dimension of the total Hilbert space. This quantity measures the non-commutativity between a Heisenberg operator $O_1(0,t)=U^\dag O_1(0)U$ and a time independent operator $O_2(r,0)$ initially separated by a distance $r$, where $U$ is the time evolution operator.   In RUC models, the growth of the Heisenberg operator $O(t)$ reduces to the solution of a classical stochastic problem:  the biased random walk \cite{nahum_operator_2018,von_keyserlingk_operator_2018}.  This directly implies that $C(r,t)\sim C((r-v_{\mathrm{B}} t)/\sqrt{t})$, where $v_{\mathrm{B}}$ is the butterfly velocity and the $1/\sqrt{t}$ factor indicates the diffusive broadening of the front. Inspired by this work, a series of RUCs with different symmetries have been proposed \cite{rakovszky_diffusive_2017, khemani_operator_2017, Rowlands2018}, which introduce extra conservation laws in the quantum dynamics and give rise to diffusive transport on top of ballistic information propagation in conventional models.  With the right conservation laws \cite{Pai2019, kn, sala}, localization is also possible.

Here, we consider both random and non-random discrete time quantum circuits with $\mathrm{U}(1)$ symmetry and explore operator growth and OTOCs in distinct symmetry sectors associated to the conserved charge.  To be precise, we consider a one dimensional chain of length $L$, with a spin-$\frac{1}{2}$ degree of freedom on every site, and a conserved $S^z$.  The Hilbert space $\mathcal{H} = (\mathbb{C}^2)^{\otimes L}$ can be written (in the obvious product state basis) as the direct sum of subspaces with a fixed number of up spins: \begin{equation}
\mathcal{H} = \bigoplus_{N^\uparrow=1}^L \mathcal{H}^{N^\uparrow},
\end{equation}with \begin{equation}
    \mathcal{H}^{N^\uparrow} = \mathrm{span}\left\lbrace |s_1s_2\cdots s_L\rangle : \sum_{i=1}^Ls_i = 2N^\uparrow-L \right\rbrace.
\end{equation}Here $s_i= 1$ or $-1$ represents whether a spin is up ($\uparrow$) or down ($\downarrow$) respectively.  Define projectors $\mathcal{P}^{N^\uparrow}$ onto subspaces $\mathcal{H}^{N^\uparrow}$,  we are interested in discrete time quantum evolution arising from a many-body unitary matrix $U(t)$ obeying \begin{equation}
    [\mathcal{P}^{N^\uparrow}, U(t)] = 0.
\end{equation} 

In particular, we will focus on the regime with small but finite density $\alpha = N^\uparrow/L$ of conserved charge in the large $L$ limit.  These are highly polarized states which are exponentially rare in $\mathcal{H}$, but are often of particular physical interest.  For example, these could serve as crude models for dynamics in low temperature states in Hamiltonian quantum systems.  We compare and contrast diffusive transport and operator growth in a RUC with a non-random, chaotic Floquet circuit (CFC).  Our results are summarized in Table \ref{tab:comparision}.

\begin{table}[t]
\begin{center}
\begin{tabular}{ | m{1.5cm}| m{1.5cm} | m{2.5cm} | m{1.5cm} | }
\hline
 $v_{\mathrm{B}}$ (RUC) & $D$ (RUC) & $v_{\mathrm{B}}$ (CFC) & $D$ (CFC)\\
\hline
\multirow{2}{*} {$\sim \alpha$} & \multirow{2}{*} {$O(1)$} &   short time: $O(1)$  & \multirow{2}{*} {$\sim 1/\alpha$} \\
& & long time: $\sim \alpha$ & \\
\hline
\end{tabular}
\end{center}
\caption{The comparison of the quantum dynamics between RUC and CFC.}
\label{tab:comparision}
\end{table}

In Section \ref{sec:ruc}, we study the RUC dynamics in a system of spin-$\frac{1}{2}$ degrees of freedom with $z$-magnetization conserved.  Analytically, we argue that operators spread analogously to a classical biased random walker.  The bias in the random walk occurs when, loosely speaking, operators that act on two spins of the same orientation collide, and so $v_{\mathrm{B}} \sim \alpha$.  The diffusion of the wave front, and of the conserved charge, is controlled by the classical stochastic noise, and $D\sim \alpha^0$ does not strongly depend on density.  We verify this prediction numerically in large scale simulations of a quantum automaton RUC \cite{Gopalakrishnan2018, Gopalakrishnan_Zakirov2018, Iaconis2019, Alba2019}.

In Section \ref{sec:cfc}, we discuss diffusion and operator growth in a CFC.  At early times, and at small $\alpha$, the dynamics is controlled by the coherent quantum walk of an effectively single particle operator, and so the diffusion constant $D\sim 1/\alpha$ and $v_{\mathrm{B}} \sim \alpha^0$.  At late times, rare multi-particle collisions begin to dephase the quantum walk and the operator wavefront spreads as $v_{\mathrm{B}}\sim \alpha$, as in the RUC.  However, we argue analytically that the diffusion constant maintains its anomalous scaling with $\alpha$ by analogy to a noisy quantum walk of a single particle.  We numerically justify our arguments about the late time dynamics using a particular model of a CFC, described below.

\section{Random circuit dynamics} \label{sec:ruc}
\begin{figure}
\centering
\includegraphics[width=.9\columnwidth]{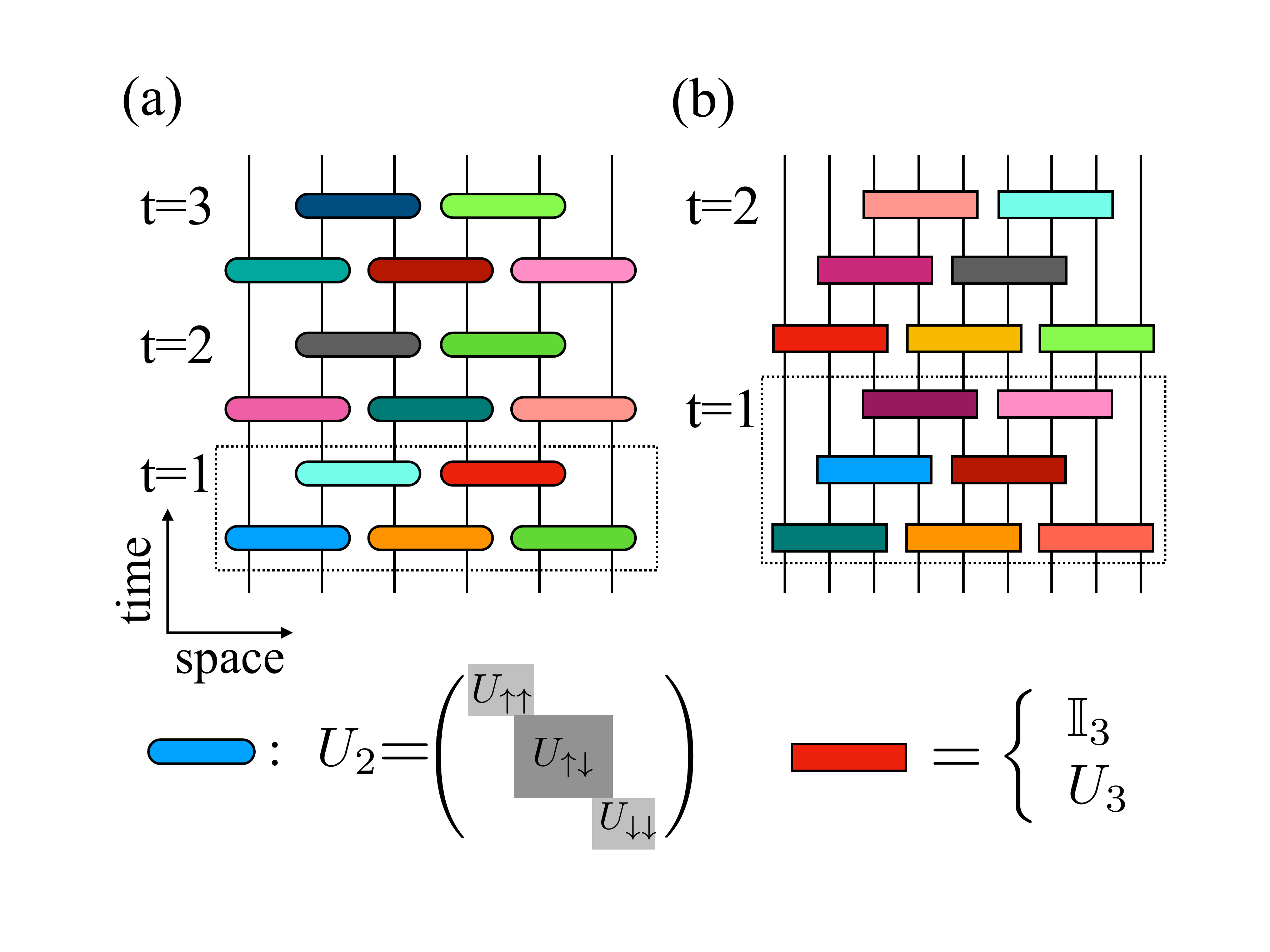}
\caption{(a) The Haar random circuit with $\mathrm{U}(1)$ symmetry. A single period of the circuit consists two layers. The rounded block is a two-qubit gate where $U_{\uparrow\uparrow}$ and $U_{\downarrow\downarrow}$ are independent random element of $\mathrm{U}(1)$ and $U_{\uparrow\downarrow}$ is a Haar random unitary element in $U(2)$. (b) The quantum automaton circuit with $\mathrm{U}(1)$ symmetry. A single period of the circuit consists three layers. The block is a three-qubit gate which randomly picks an identity operator or Fredkin gate. The dashed boxes in both (a) and (b) indicate the circuit in one time step.}
\label{fig:random_circuit}
\end{figure}

In a RUC, the time evolution operator $U(t)$ is produced by a quantum circuit which is composed of the staggered layers of (products of) a random two-qubit gate $U_2(r,t)$: 

\begin{align}
    U(t) &= \prod_{m=1}^t \left(\prod_{j=2,4,\ldots} U_2(j,m)\right) \notag \\
    &\times\left(\prod_{j=1,3,\ldots} U_2(j,m)\right) \label{eq:RUCdef}
\end{align} where $U_2(j,m)$ represents a Haar random two-qubit gate acting on the spins at sites $j$ and $j+1$  at time $m$, subject to the constraint \begin{equation}
    [\mathcal{P}^{N_{\uparrow}}, U_2(j,t)] = 0, \;\;\; (\text{for any }N_{\uparrow}).
\end{equation}
See Fig.~\ref{fig:random_circuit} (a).  
Previous calculations of operator dynamics have largely studied OTOCs in the entire ensemble, as defined in (\ref{eq:otocintro}) \cite{khemani_operator_2017,rakovszky_diffusive_2017}, although the dependence of butterfly velocity on chemical potential was examined in \cite{rakovszky_diffusive_2017}.  In the thermodynamic limit, operator dynamics in the entire ensemble will be dominated by the dynamics at density $\alpha=N_\uparrow/L\approx 1/2$. In this paper, we are more interested in sectors with small $\alpha \ll 1$, where we will show that $\mathcal{P}$ plays a significant role.

\subsection{Full polarization}

Consider the OTOC in a polarized sector 
\begin{align}
C_{\mathrm{XX}}(r,t)=-\frac{\mbox{Tr}\left\{\left( \mathcal{P}^{N_{\uparrow}}\left[X_x(t), X_{x+r}(0)\right]\right)^2\right\}}{\mbox{Tr}\mathcal{P}^{N_{\uparrow}}}
\label{eq:OTOC}
\end{align}
where $X(t)=U(t)^\dag X U(t)$ is a Heisenberg time evolved local operator.  Here the $X$ operator denotes the Pauli X matrix.

\begin{figure*}[t]
\centering
 \subfigure[]{\label{fig:Haar_N_0_collapse} \includegraphics[width=.67\columnwidth]{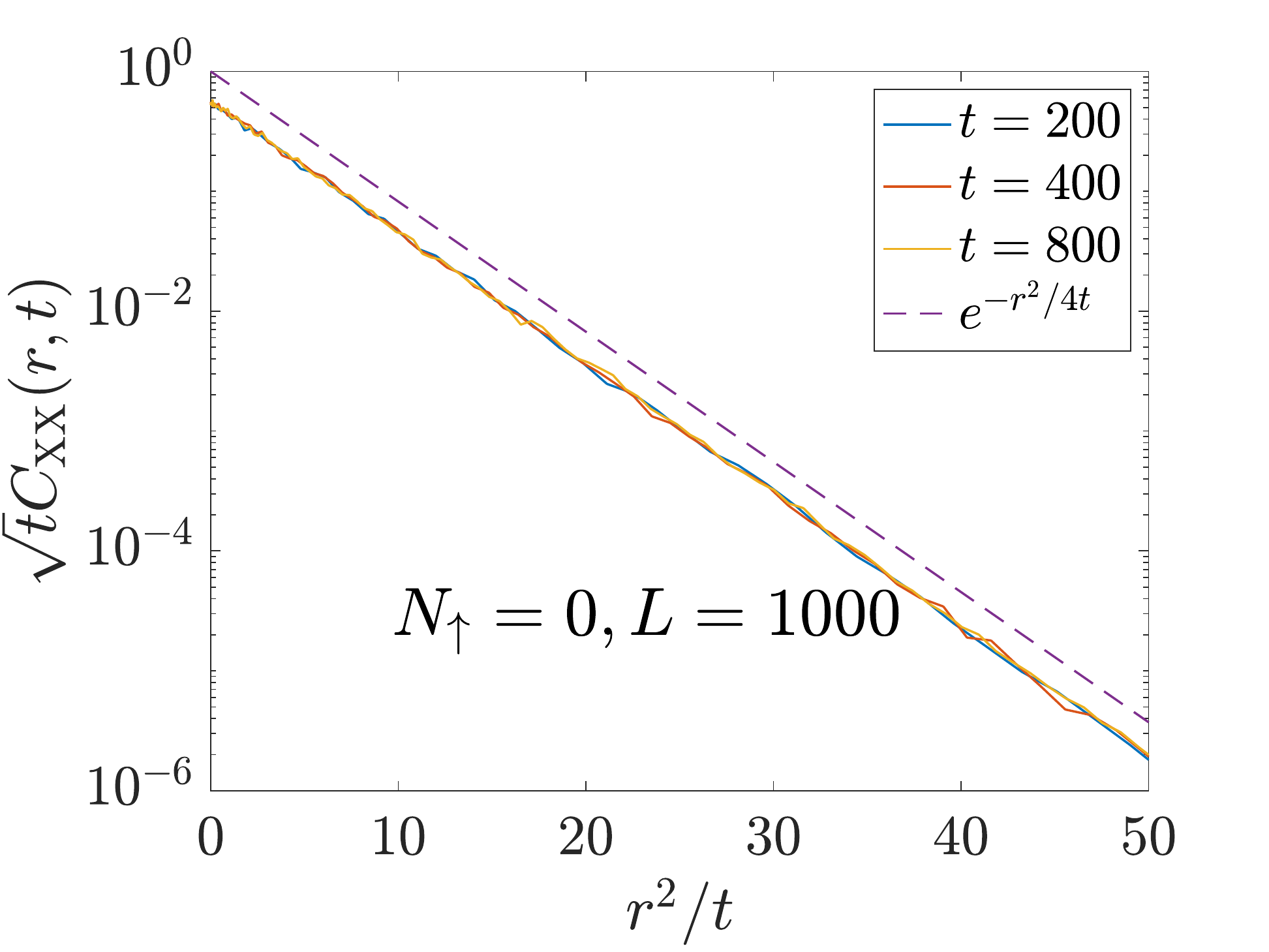}}
 \subfigure[]{\label{fig:Haar_N_1_collapse} \includegraphics[width=.67\columnwidth]{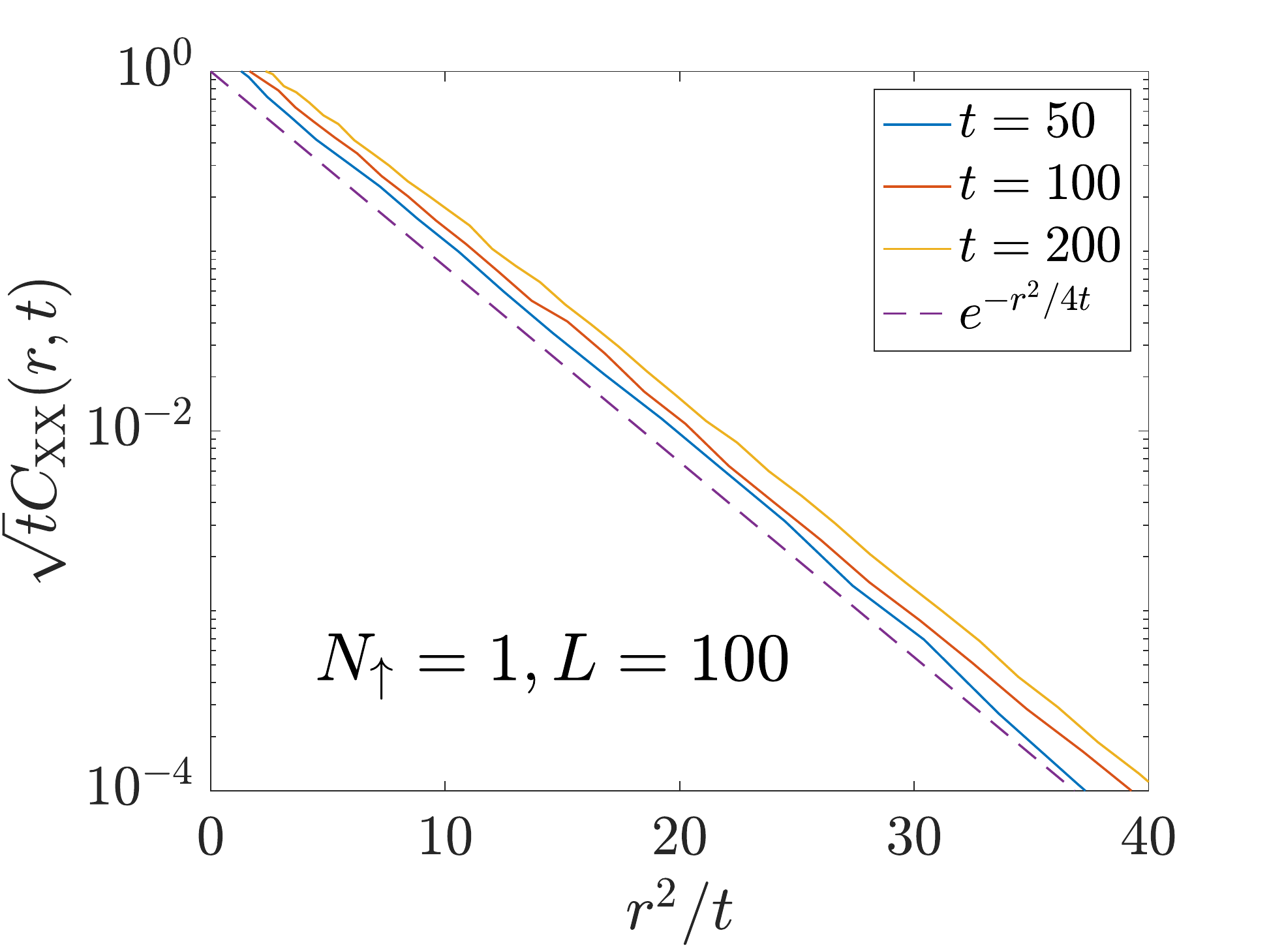}}
  \subfigure[]{\label{fig:Haar_N_2_collapse} \includegraphics[width=.67\columnwidth]{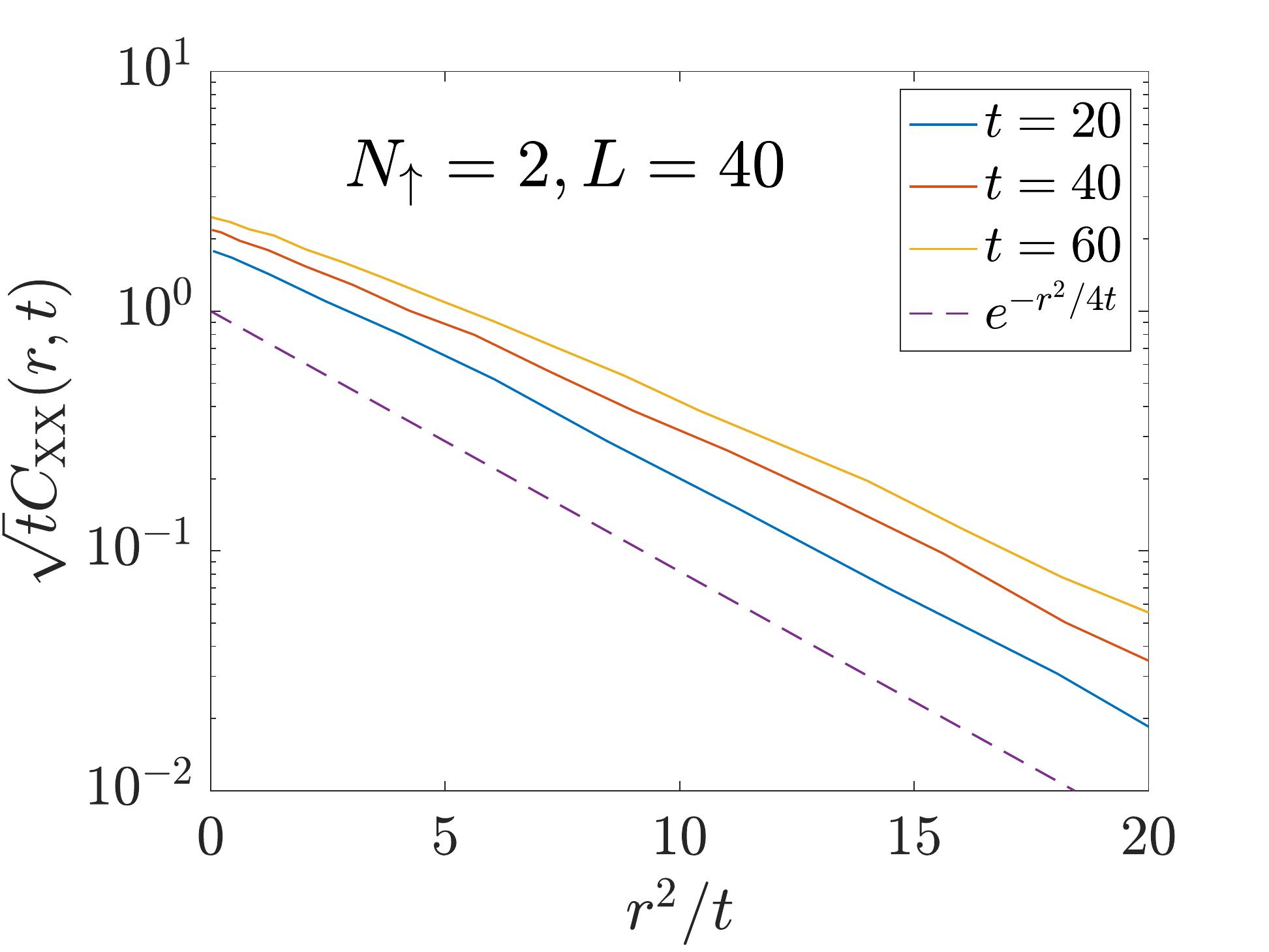}}
\caption{ (a) The data collapse of $C_{\rm XX}(r,t)$ in the full polarized sector with $N_{\uparrow}=0$. (b)  $C_{\rm XX}(r,t)$ vs $r^2/t$ in the $N_{\uparrow}=1$ sector. (c) $C_{\rm XX}(r,t)$ vs $r^2/t$ in the $N_{\uparrow}=2$ sector. In all these three plots, $C_{\rm XX}(r,t)$ is obtained by taking average over more than 200 realizations.} 
\label{fig:OTOC_haar}
\end{figure*}

First, consider the fully polarized sector with $\alpha=N^\uparrow =0$:\begin{equation}
    \mathcal{P}^0=|\downarrow\downarrow\ldots\rangle\langle \downarrow\downarrow\ldots| = |0\rangle\langle 0|.
\end{equation}
In this simple limit, 
\begin{align}
C_{\mathrm{XX}}(r,t)&=4(\mbox{Im}\langle 0|X^-_{x+r}(0)X^+_x(t)|0\rangle)^2\nonumber\\
&=4(\mbox{Im}\langle 0|X^-_{x+r}U^\dag (t) X^+_x|0\rangle)^2 \label{eq:CXXalpha0},
\end{align} 
where $X^\pm=(X\pm \mathrm{i} Y)/2$.  If $|L/2-x| \ll L$, the OTOC is approximately independent of $x$ at short times.  Observe that 
\begin{equation}
    X^+_x|0\rangle=|\downarrow_{1}\ldots \downarrow_{x-1}\uparrow_x\downarrow_{x+1}\ldots\downarrow_{L}\rangle.
\end{equation}
Under the unitary time evolution $U^\dag (t)$, the up spin at $x$ will move to other sites; by charge conservation, there will always be exactly one up spin in $U^\dag(t)X^+_x(0)|0\rangle$.   The probability that the up spin is on site $x+r$ at time $t$ is $|\langle 0| X^-_{x+r}U^\dag (t)X^+_x|0\rangle |^2$, which is quite similar to (\ref{eq:CXXalpha0}).  Applying the $U^\dag(t)$ operator given by (\ref{eq:RUCdef})  and taking average over different circuit realizations, it is easy to see that this solitary up spin is performing a single particle classical random walk in the background of down spins.  The diffusion constant $D=1$ for this classical random walk is derived in Appendix \ref{app:1partD}.  In the continuum limit, $C(r,t)$ becomes a Gaussian distribution:
\begin{equation}
    C(r,t)\sim \frac{\mathrm{e}^{-r^2/4t}}{\sqrt{t}}.
    \label{eq:C_diffusive}
\end{equation}
Quantum information spreads diffusively with butterfly velocity $v_{\mathrm{B}}=0$. We numerically confirm this result in Fig.~\ref{fig:Haar_N_0_collapse}.  

It is instructive to also discuss this single particle quantum walk directly in the language of growing operators.  We define a single site basis for Hermitian operators:
\begin{align}
\{ P^{\uparrow,\downarrow}=(I\pm Z)/2, X^{\pm}=(X\pm \mathrm{i}Y)/2 \}.
\label{eq: op_basis}
\end{align} 
The space of many-body Hermitian operators is a tensor product of this local basis. Any operator can be written as a superposition of these basis operators (Pauli string operators). Compared with the conventional $\otimes\{ X, Y, Z, I\}$ basis, this new basis is more convenient when there is $\mathrm{U}(1)$ symmetry. 

In order to analyze the OTOC in the polarized sector, we need to study the dynamics of the projected operator $X^+_x(t) \mathcal{P}^{N_\uparrow}$.  For the fully polarized sector with $N_\uparrow=0$, defining the spin propagator $G_{xy}(t)$, spin conservation demands that \begin{align}
    X^+_x(t) \mathcal{P}^{0} &= \sum_{y=1}^L G_{x,y}(t) X^+_y\mathcal{P}^0 \notag \\
    &= \sum_{y=1}^L G_{x,y}(t) P^\downarrow_1 \cdots P^\downarrow_{y-1} X^+_y P^\downarrow_{y+1}\cdots P^\downarrow_{L},
\end{align} 
where $\sum_{y}|G_{x,y}(t)|^2=1$.
The result in \eqref{eq:C_diffusive} can easily be understood by observing that  $P^\downarrow P^\downarrow$ is invariant under $U_2$, while \begin{align}
    \mathbb{E}\left[U_2^\dagger X^+_1 P^\downarrow_2 U_2 \right] &=\mathbb{E}\left[U_2^\dagger X^+_2 P^\downarrow_1 U_2 \right] \notag \\  &= \frac{X^+_1 P^\downarrow_2+X^+_2 P^\downarrow_1 }{2},
\end{align}
where $\mathbb{E}[\cdot]$ denotes average over different circuit realizations. In other words, in the many-body operator $X^+_x(t) \mathcal{P}^{N_\uparrow}$, the lone $X^+$ performs a classical random walk in a background of $P^{\downarrow}$ and the averaged distribution function $\mathbb{E}\left[|G_{x,x+r}|^2\right]$ satisfies the Gaussian distribution in \eqref{eq:C_diffusive}. The operator growth is entirely characterized by the location of the $X^+$ operator, for which the distribution function spreads out diffusively as time evolves. 


\subsection{Finite polarization}

Now we turn to the dynamics of operator growth when $\alpha>0$.  It is most convenient to work in the operator language and study $X^+_x(t) \mathcal{P}^{N_\uparrow}$ at finite but small density $\alpha$.   The allowed transition rates governed by the $U_2$ can be summarized as follows: the following (unnormalized) operators mix into a random superposition of the others in the same set: see Appendix \ref{app:rucrules}. 
For a growing operator $X^+_x(t)\mathcal{P}^{N_\uparrow}$:  (1) the total number of $P^{\uparrow}$ and $X^+$ is equal to $N_{\uparrow}+1$; (2) the number of $X^+$ is always larger than $X^-$ by one.

We now try to estimate the location of the right most $X^+$ operator in $X^+_x(t)\mathcal{P}^{N_\uparrow}$, keeping in mind that (most likely) this location serves as a reasonable proxy for the right\footnote{The left edge obeys an analogous description.} ``edge" of the growing operator in the restricted ensemble.  Let us place an auxiliary hat label on this right most $X^+$:  $\widehat{X}^+$. If $\widehat{X}^+$ operator has neighbor $P^\uparrow$ or $P^\downarrow$, it performs an unbiased random walk, as in the $N_{\uparrow}=0$ sector.  If the neighbor is $X^-$, then it is possible that $\widehat{X}^+$ is destroyed and turned into a $P^{\uparrow/\downarrow}$; however, we expect that in this case, with high probability an $X^+$ will re-emerge at a similar location.  As such, we will assume that the presence of $X^-$ do not qualitatively alter the dynamics of $\widehat{X}^+$.  The remaining possibility is that $\widehat{X}^+$ has another $X^+$ to its left, in which case the $\widehat{X}^+$ is blocked from moving to the left.  This blocking of the random walking $\widehat{X}^+$ imparts some bias into the motion, which will cause $\widehat{X}^+$ to tend to drift to the right.  Since the density of $X^+$ and $P^\uparrow$ should scale as $\alpha$, we expect that with probability $\sim \alpha$ we encounter an $X^+$ which biases the random walk, leading to the estimate  \begin{equation}
    v_{\mathrm{B}}\sim \alpha.  \label{eq:vBalpha}
\end{equation}
We also note that in this description, the diffusion constant of the conserved charge, and of the operator wave front, is dominated by the single particle random walk described previously; hence, \begin{equation}
    D \sim \alpha^0
\end{equation}
does not strongly depend on polarization.\footnote{This result contradicts \cite{rakovszky_diffusive_2017} who claimed that the diffusion constant diverges in the limit of full polarization. Ref. \cite{rakovszky_diffusive_2017} estimated the diffusion constant from the correlator $\langle Z_r(t) Z_r(0)\rangle - \langle Z_r\rangle^2$. They found this correlator vanished in the limit of  full polarization, and since this correlator is proportional to $(Dt)^{-1/2}$ they concluded that the diffusion constant must diverge. However, close to  full polarization even the $t\rightarrow 0$ limit of this correlator vanishes, and the vanishing of the correlator observed simply reflects this $t=0$ normalization, and not any divergence of the diffusion constant. }

For this argument to be correct, it is necessary that the right-moving $X^+$ ``scrambles" the projectors in its wake:  the number of $X^+$ in the operator should scale as $\sim \alpha t$.  We expect that this does indeed occur due to random dephasing of $P^\uparrow$ and $P^\downarrow$ as $X^+$ moves through -- this can spawn pairs of $X^+$ and $X^-$ according to the transition rules above.  As such, we expect that it is exponentially unlikely to find $\widehat{X}^+$ a distance $\gg \alpha^{-1}$ away from another $X^+$.



To confirm the cartoon above, we numerically compute operator growth in the Haar RUC with $\mathrm{U}(1)$ symmetry with small $N_{\uparrow}$. Due to the constraint that the superposition $P_1^{\downarrow}P_2^{\uparrow}+P_1^{\uparrow}P_2^{\downarrow}$ is invariant under $U_2$ gate, the Haar  RUC with $\mathrm{U}(1)$ symmetry cannot be mapped to a simple  Markov chain, and therefore we cannot perform classical simulation for large system size. We directly simulate this quantum RUC for small system size. The result is presented in  Fig.~\ref{fig:Haar_N_1_collapse} and Fig.~\ref{fig:Haar_N_2_collapse}. We observe that the curves do not collapse into a single curve if we employ the previous scaling form for $N_{\uparrow}=0$ sector. 

\subsection{Quantum automaton circuit}

Of course, this does not demonstrate the emergence of a finite butterfly velocity.  To observe this behavior reliably in our numerics, we now turn to a different RUC: the quantum automaton (QA) \cite{Gopalakrishnan2018, Gopalakrishnan_Zakirov2018, Iaconis2019, Alba2019}, which we expect exhibits similar operator growth to the Haar RUC, while allowing for large scale numerical simulation.

As shown in Fig.~\ref{fig:random_circuit} (b), our QA consists of three-qubit gates, which are randomly chosen to be $U_3$ with probability $f$ or the identity with probability $1-f$. We choose $U_3$ to be the Fredkin gate \cite{Fredkin1982},
\begin{align}
U_3\equiv (1-Q)+Q(|\downarrow\downarrow\uparrow\rangle\langle \uparrow\downarrow\downarrow|+ |\uparrow\downarrow\downarrow\rangle\langle \downarrow\downarrow\uparrow| )
\end{align}
where the projector $Q$ is \begin{equation}
    Q=|\downarrow\downarrow\uparrow\rangle\langle \downarrow\downarrow\uparrow|+|\uparrow\downarrow\downarrow\rangle\langle \uparrow\downarrow\downarrow|
\end{equation}
Note that $U_3$ is $\mathrm{U}(1)$ symmetric: it swaps between the two states $|\downarrow\downarrow\uparrow\rangle$ and $|\uparrow\downarrow\downarrow\rangle$ and leaves other states invariant. 


Under the adjoint action of $U_3$, a Pauli string of  $P^{\uparrow,\downarrow}$ and $X^\pm$ either remains invariant or becomes another Pauli string: see Appendix \ref{app:automaton}.   This property is similar to the Clifford circuit dynamics where there is no superposition for the operator dynamics \cite{Gottesman1998}. Notice that this property only holds in the special basis defined in \eqref{eq: op_basis} and is not true in the conventional $\otimes\{X,Y,Z,I\}$ basis. For $\widehat{X}^+$ in the background of $P^{\uparrow/\downarrow}$, under $U_3$ gate, we have $\widehat{X}^+ P^{\downarrow} P^{\uparrow,\downarrow} \longleftrightarrow P^{\uparrow,\downarrow} P^\downarrow \widehat{X}^+$, therefore $\widehat{X}^+$ can move freely to the right or left with the same displacement. However, for the Pauli string $X^+X^+\widehat{X}^+$ and $X^+P^{\uparrow}\widehat{X}^+$, the location of $\widehat{X}^+$ operator is invariant under $U_3$. This gives rise to a biased random walk for $\widehat{X}^+$.  The same logic as before leads to the prediction (\ref{eq:vBalpha}).   

\begin{figure}[hbt]
\centering
 \subfigure[]{\label{fig:a_01} \includegraphics[width=.8\columnwidth]{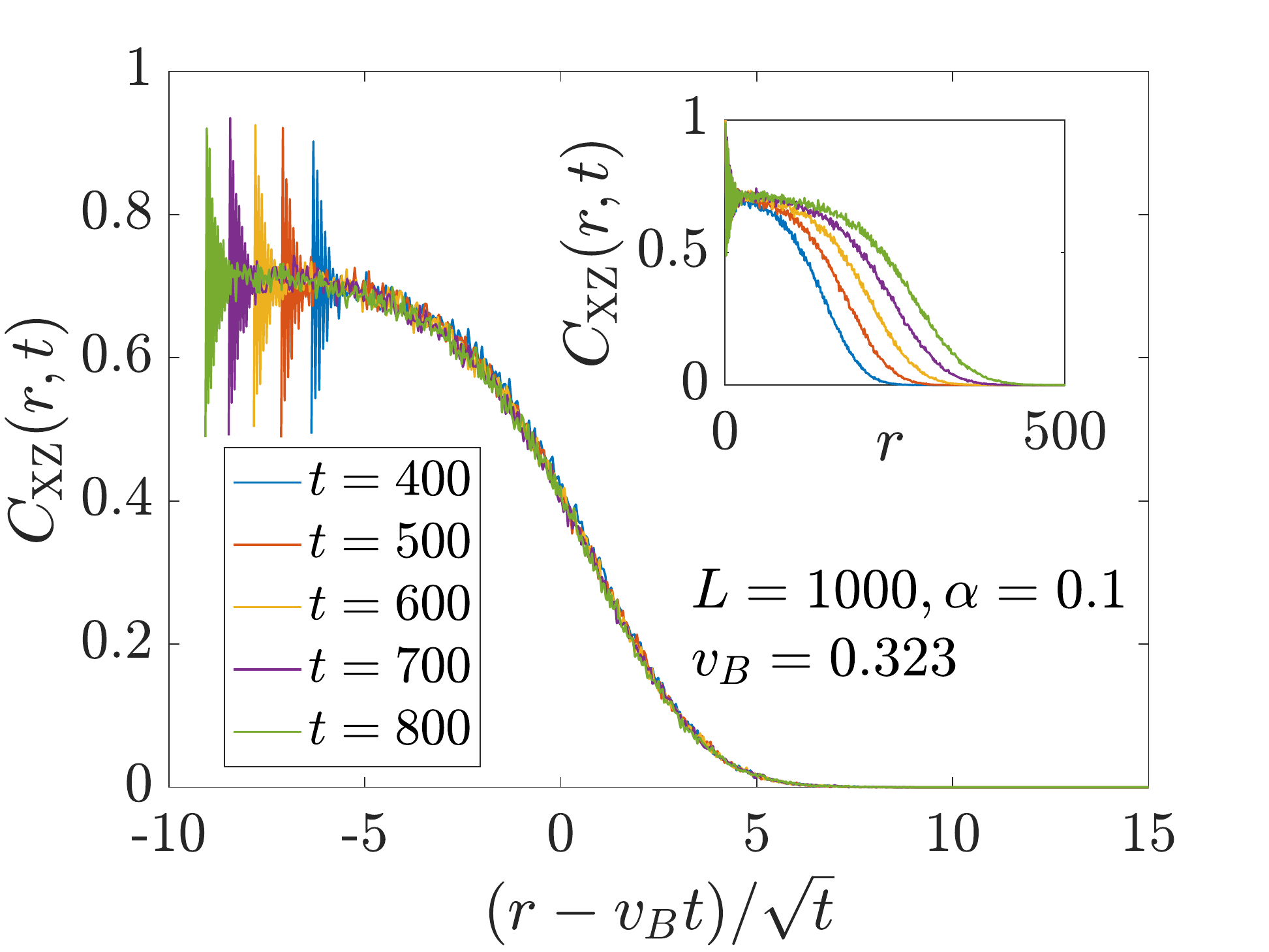}}
 \subfigure[]{\label{fig:vB_a} \includegraphics[width=.8\columnwidth]{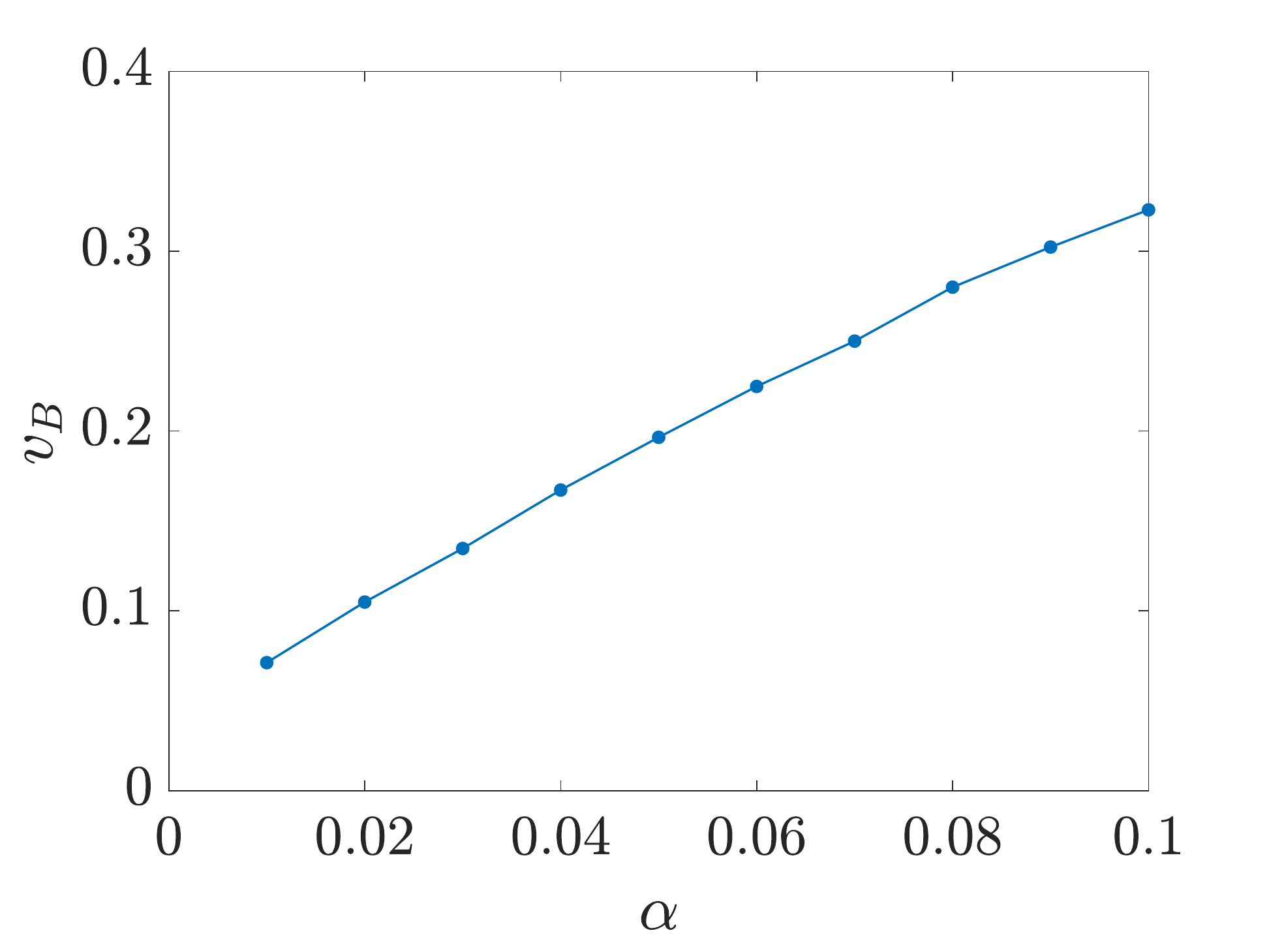}}
\caption{ (a) The data collapse for the front of OTOC in QA circuit. The strong oscillatory behavior observed in the short distance regime is due to a special feature of the three-qubit Fredkin gate, which can only swap the spin configuration between the first and third qubits. (b) The butterfly velocity vs $\alpha$. In both plots, we take the probability $f=0.5$.} 
\label{fig:OTOC}
\end{figure}

Since Pauli strings map to Pauli strings, we can easily perform large scale classical simulations of the QA, in contrast to the Haar RUC.  For numerical ease, we study a slightly different OTOC:
\begin{align}
C_{\rm XZ}(r,t)&=-\frac{\mbox{Tr}\left\{ \mathcal{P}^{N_{\uparrow}}\left[ X_x(t), Z_{x+r}\right]^2\right\}}{\mbox{Tr}\mathcal{P}^{N_{\uparrow}}}\nonumber\\
&=\sum_{s,s^\prime}\frac{\left|\langle s| [X_x, Z_{x+r}(-t)|s^\prime]\rangle\right|^2}{\mbox{Tr}\mathcal{P}^{N_{\uparrow}}}\nonumber\\
&=\sum_s\frac{\left| \langle s|Z_{x+r}(-t)|s\rangle-\langle s^*|Z_{x+r}(-t)|s^*\rangle \right|^2}{\mbox{Tr}\mathcal{P}^{N_{\uparrow}}}
\end{align}
where $|s^*\rangle=X_x|s\rangle$. In Fig.~\ref{fig:OTOC}, we present the numerical results for $C_{\rm XZ}$.   We numerically find that the front of OTOC is captured by a scaling function
\begin{equation}
C_{\mathrm{XZ}}(r,t) = F\left(\frac{r-v_{\mathrm{B}}t}{\sqrt{t}}\right)
\end{equation}
for a function $F$ which may depend on $\alpha$.  Physically, this $F$ reveals a ballistically propagating wave front with diffusive broadening, in agreement with earlier work \cite{rakovszky_diffusive_2017,khemani_operator_2017}.  When the density $\alpha$ is small, we further find (\ref{eq:vBalpha}) numerically, consistent with our analytical arguments.
\section{Floquet non-random dynamics}\label{sec:cfc}
For a realistic chaotic system without randomness, it is believed that at high temperature, or in the polarized sector with $\alpha\approx 1/2$, the quantum dynamics can be well approximated by the RUC models with the same symmetry.   We now demonstrate that this assumption generically fails when $\alpha \ll 1$.  

\begin{figure}
\centering
 \subfigure[]{\label{fig:Floq_circuit} \includegraphics[width=.4\columnwidth]{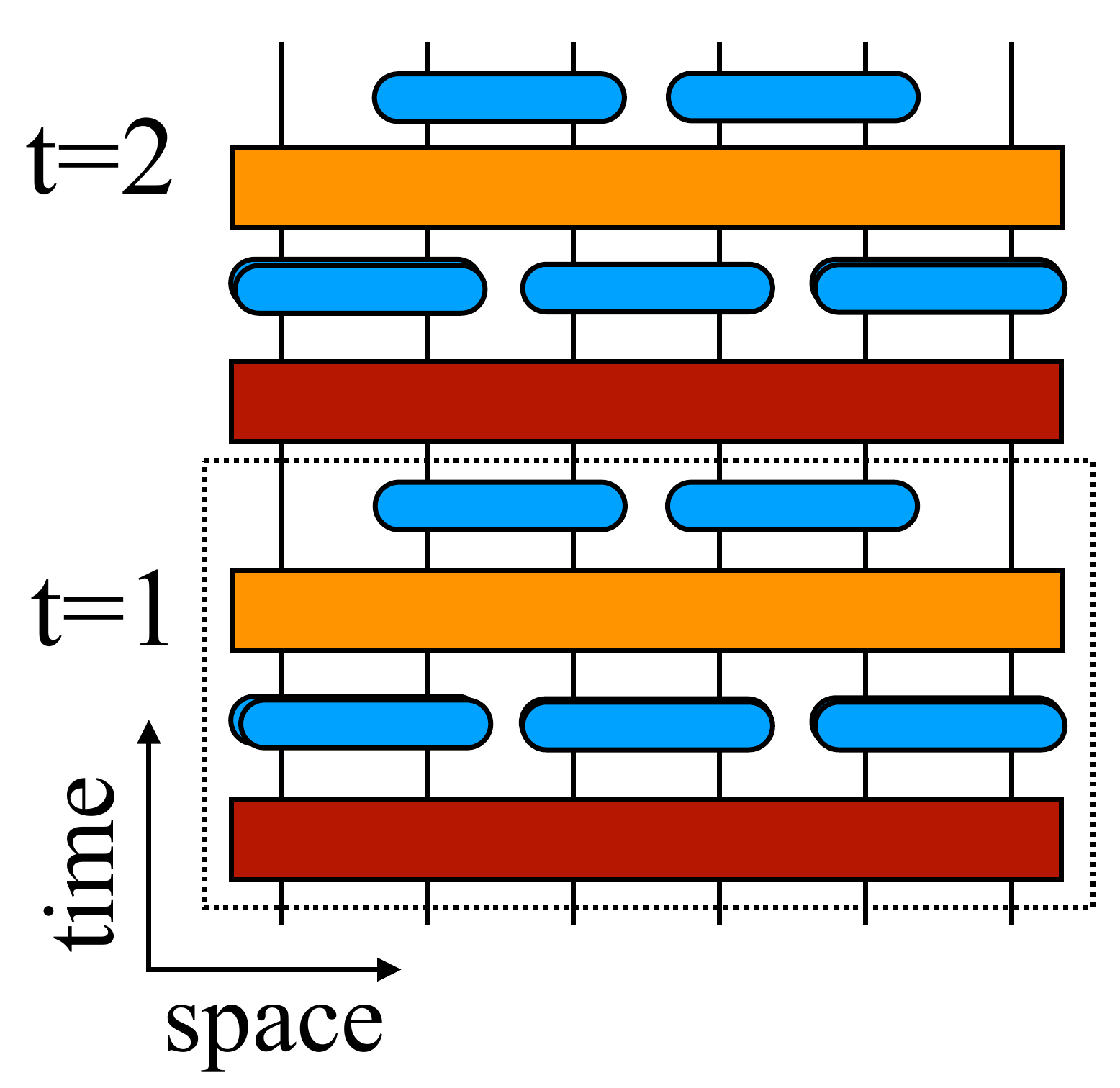}}
 \subfigure[]{\label{fig:quantum_walk} \includegraphics[width=.56\columnwidth]{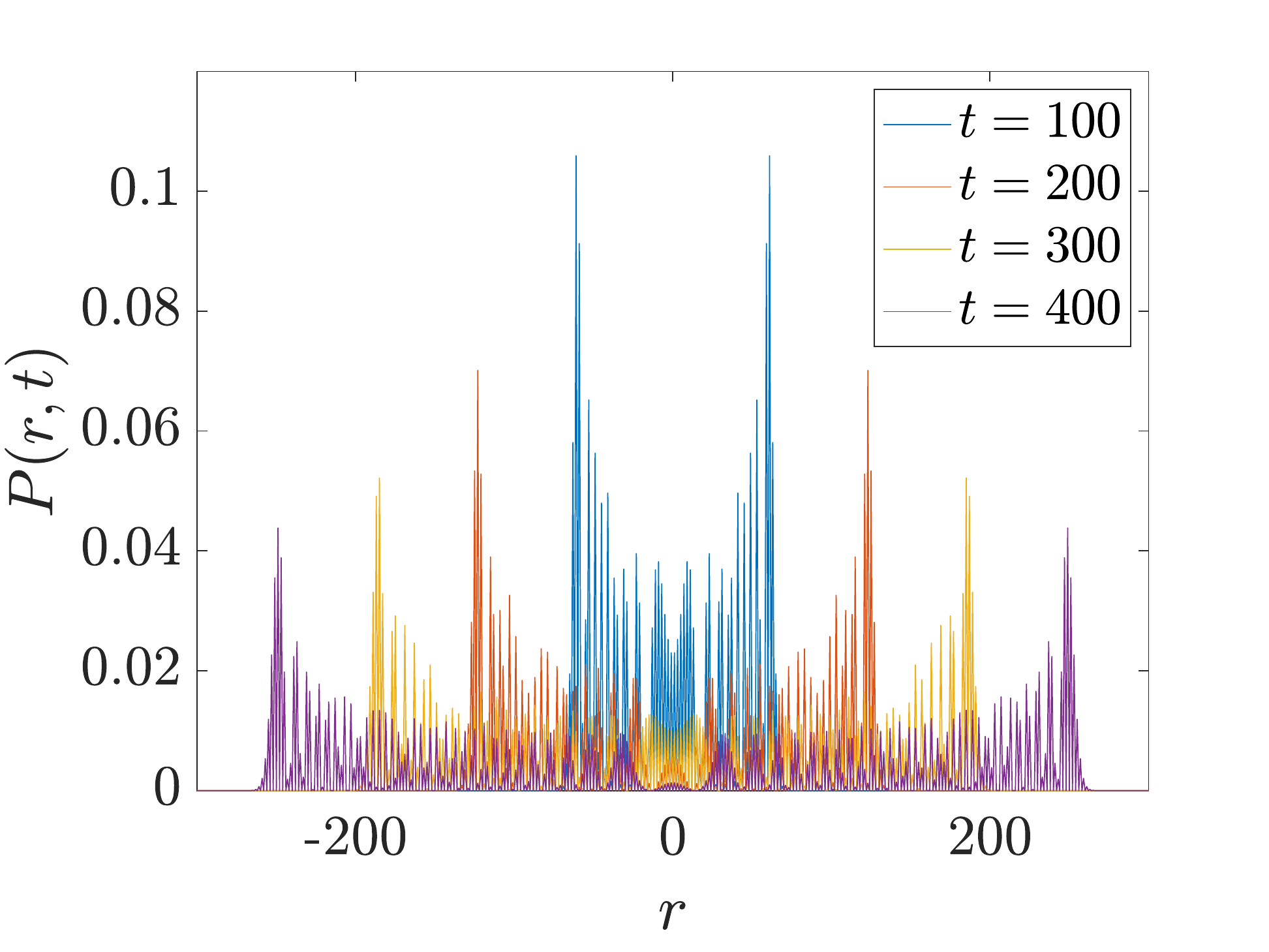}}
\caption{ (a) The schematics for the Floquet circuit. (b) The probability distribution for the up spin under time evolution generated with the staggered $U_{m,m+1}$ gate.  In order to obtain a symmetric quantum walk, we take the initial state as $1/\sqrt{2}(|0\rangle+i|1\rangle)$. The probability distribution $P(r,t)$ can be considered as $C(r,t)$ for $X$ operators in the $N_{\uparrow}=0$ sector.} 
\label{fig:CFC}
\end{figure}

We numerically study a chaotic Floquet circuit (CFC), defined as follows: for integer times $t=0,1,\ldots$, \begin{equation}
    U(t) = U_{\mathrm{F}}^t
\end{equation}
where as shown in Fig.~\ref{fig:Floq_circuit},
\begin{align}
U_{\mathrm{F}}=\left(\prod_m U_{2m,2m+1}\right)  U_{Z,2}\left(\prod_m U_{2m-1,2m}\right) U_{Z,1}
\end{align}
where $U_{2m,2m+1}$ and $U_{2m-1,2m}$ are defined on two neighboring qubits and preserve $\mathrm{U}(1)$ symmetry: e.g.\footnote{The choice of irrational phase factors was for convenience and is not essential to the model.} \begin{subequations}
\begin{align}
    U_{2m-1,2m}&=\exp[-\mathrm{i}\sqrt{2} (X_{2m-1}X_{2m}+Y_{2m-1}Y_{2m})], \\
    U_{2m,2m+1}&=\exp[-\mathrm{i} \sqrt{3} (X_{2m+1}X_{2m}+Y_{2m+1}Y_{2m})].
\end{align}\end{subequations}
$U_{Z,1}$ and $U_{Z,2}$ are both phase gates: \begin{subequations}
\begin{align}
    U_{Z,1} &=\prod_m \exp[-\mathrm{i} J_z Z_m Z_{m+1}], \\
    U_{Z,2} &= \prod_m \exp[-\mathrm{i} J_z Z_m Z_{m+1}Z_{m+2}].
\end{align}
\end{subequations}
The phase gates are chosen so that a product state picks up a relative phase whenever an up/down spin are adjacent.


When $J_z=0$ (corresponding to the Floquet circuit without $U_{Z,1/2}$ gates), the above model is integrable: the dynamics can be mapped to free fermions. When $J_z$ is nonzero, $U_{Z,1/2}$ will introduce extra phases to the many-body wave function, and this makes the dynamics chaotic.   However, if we consider the dynamics in a polarized sector, the non-trivial phases from the $U_{Z,1/2}$ gates are only relevant when two up spins are either nearest or next-nearest neighbors.  So even if the phases from $U_{Z,1/2}$ eventually lead to chaos, the approach to chaos may be qualitatively distinct from the RUC.


\subsection{Zero density}
We first investigate the fully polarized limit $\alpha=0$, evaluating (\ref{eq:OTOC}).  As in the RUC, this OTOC can be interpreted as follows: one of the operators $X_{x+r}(0)$ flips a single spin up, and $X_x(t)$ measures if the flipped spin is at site $x$; hence, $C_{\mathrm{XX}}(t)$ is closely related to the probability for the up spin to reach site $x$ at time $t$.   Observe that when $\alpha=0$, the phase gate $U_{Z,1/2}$ will only generate an unimportant overall phase for the entire wave function; the dynamics is only determined by the $U_{m,m+1}$ gates. By consecutively applying the $U_{m,m+1}$ gates, we see that the up spin is performing a discrete quantum walk on a line \cite{Kempe2003}.  One important feature of the quantum walk is that the dynamics is ballistic, and the quantum wave packet has a significant spread (variance $\sim t^2$): see Appendix \ref{app:qw}.  This is depicted in Fig.~\ref{fig:quantum_walk}, where we show that the ``front" of the growing operator moves with a constant velocity. This is in stark contrast with the slow dynamics observed in RUC, where the quantum coherence is completely lost and the up spin performs a classical random walk with $v_{\mathrm{B}}=0$ and variance $\sim t$.  

Alternatively, we can also interpret the above dynamics in terms of operator growth.  The projected operator $X^+(0,t)\mathcal{P}^0$ evolves as \begin{equation}
    X^+_x(t)\mathcal{P}^0 = \sum_y G_{xy}(t) X^+_y \mathcal{P}^0,
\end{equation} 
and the coefficients $G_{xy}(t)$ come from the single particle quantum walk described above.





\subsection{Low but finite density}

We now move slightly away from the $\alpha=0$ limit, and explore the highly polarized sector with small but finite $\alpha$.  The $U_{Z,1/2}$ phase gates can no longer be neglected. As two up spins quantum walk collide, the wave function will accumulate overall phases which destructively add, depending on the precise history of the walkers.  The accumulation of these phases will spoil the quantum coherence observed in the fully polarized sector with $\alpha=0$.  For example, let us consider $X^+_x(t)\mathcal{P}^{N_{\uparrow}}$.  Generically, there will be  multiple $X^+$ in the Pauli string.  When two $X^+$ operators in the background of $P^\downarrow$ are far away from each other, they are independent quantum walkers, as in the fully polarized sector.  However, collisions of two $X^+$ operators dephase the quantum walks due to the $U_{Z,1/2}$ gates.  In addition, neighboring $P^\uparrow$ and $X^-$ in the Pauli string $X^+_x(t)\mathcal{P}^{N_{\uparrow}}$ also contribute to dephasing.  Our goal is now to determine the dephasing time of this many-body operator, together with the consequences on operator growth and transport. 

%


Numerically computing the OTOC and analyzing the long time dynamics is hard in the CFC.  It is slightly simpler to numerically study transport physics, such as the relaxation of a local perturbation. If decoherence qualitatively changes the growth of operators, we can observe a transition from ballistic (quantum walk) to diffusive (classical walk) transport.   We start with a random state $|\psi\rangle$ uniformly drawn from the Hilbert space $\mathcal{H}^{N_\uparrow -1}$.
Note that this state has $\langle Z_y\rangle = 2\alpha-1$ for every site $y$.  Then we consider the initial state \begin{equation}
    |\psi_0\rangle = X^+_x |\psi\rangle,
    \label{eq:flip_up}
\end{equation} which generates some inhomogeneity in $\langle Z_y\rangle$.  We then numerically evolve the wave function forward in time, and track the relaxation of $\langle Z_y(t)\rangle$.

\begin{figure*}
\centering
\subfigure[]{\label{fig:L_160_Jz_1} \includegraphics[width=.8\columnwidth]{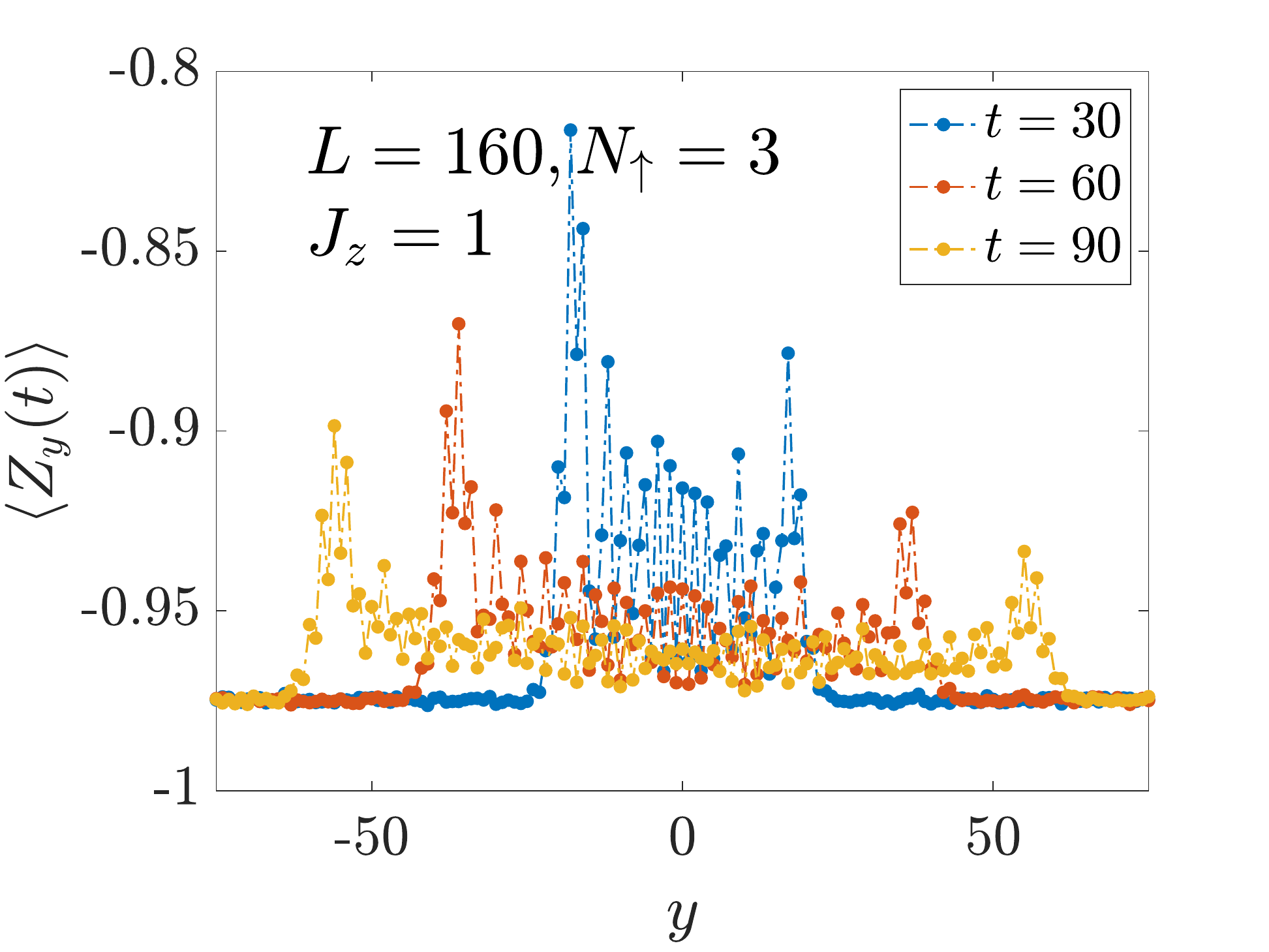}}
\subfigure[]{\label{fig:peak_t} \includegraphics[width=.8\columnwidth]{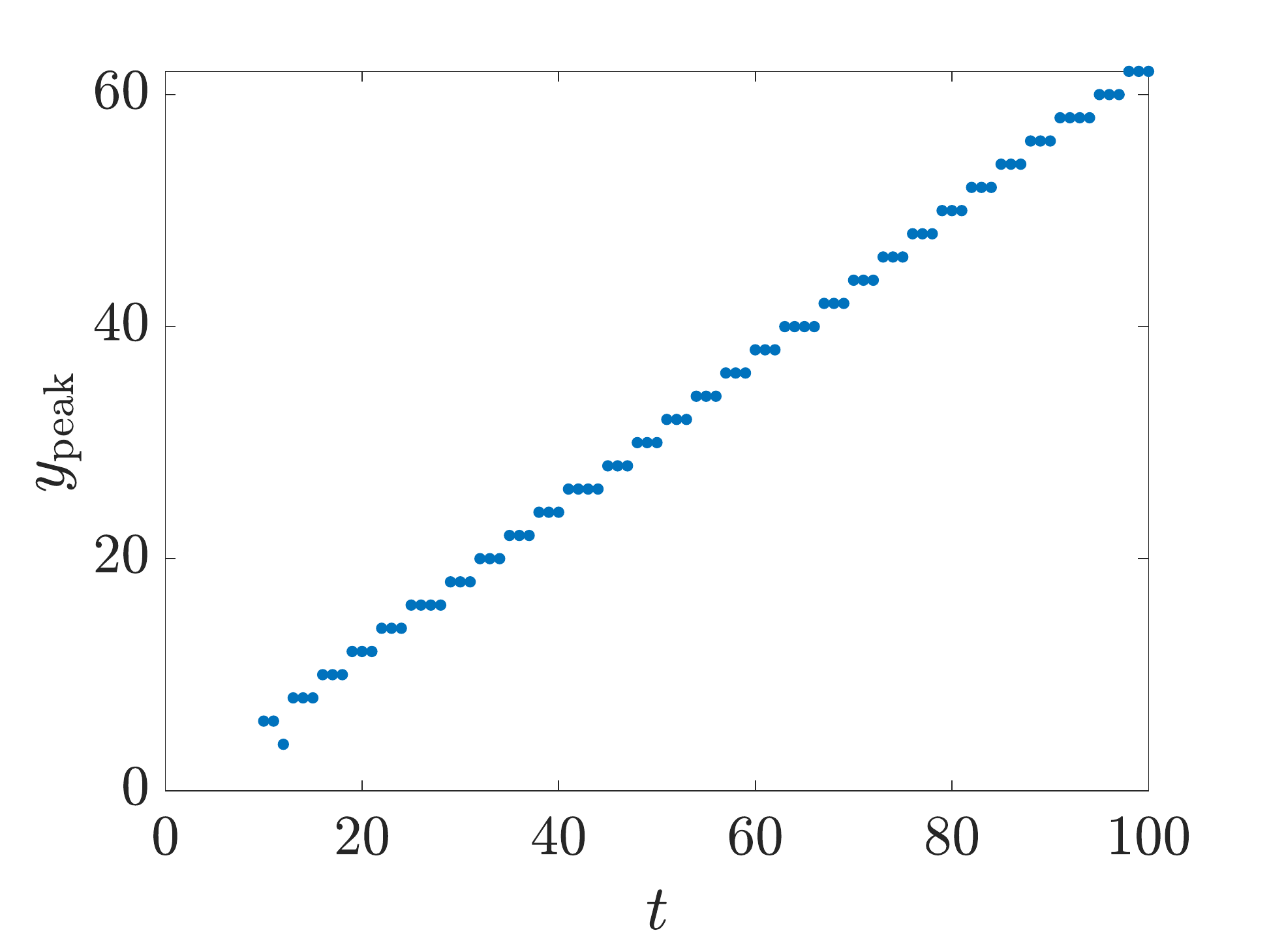}}
\subfigure[]{\label{fig:L_36_Jz_1} \includegraphics[width=.8\columnwidth]{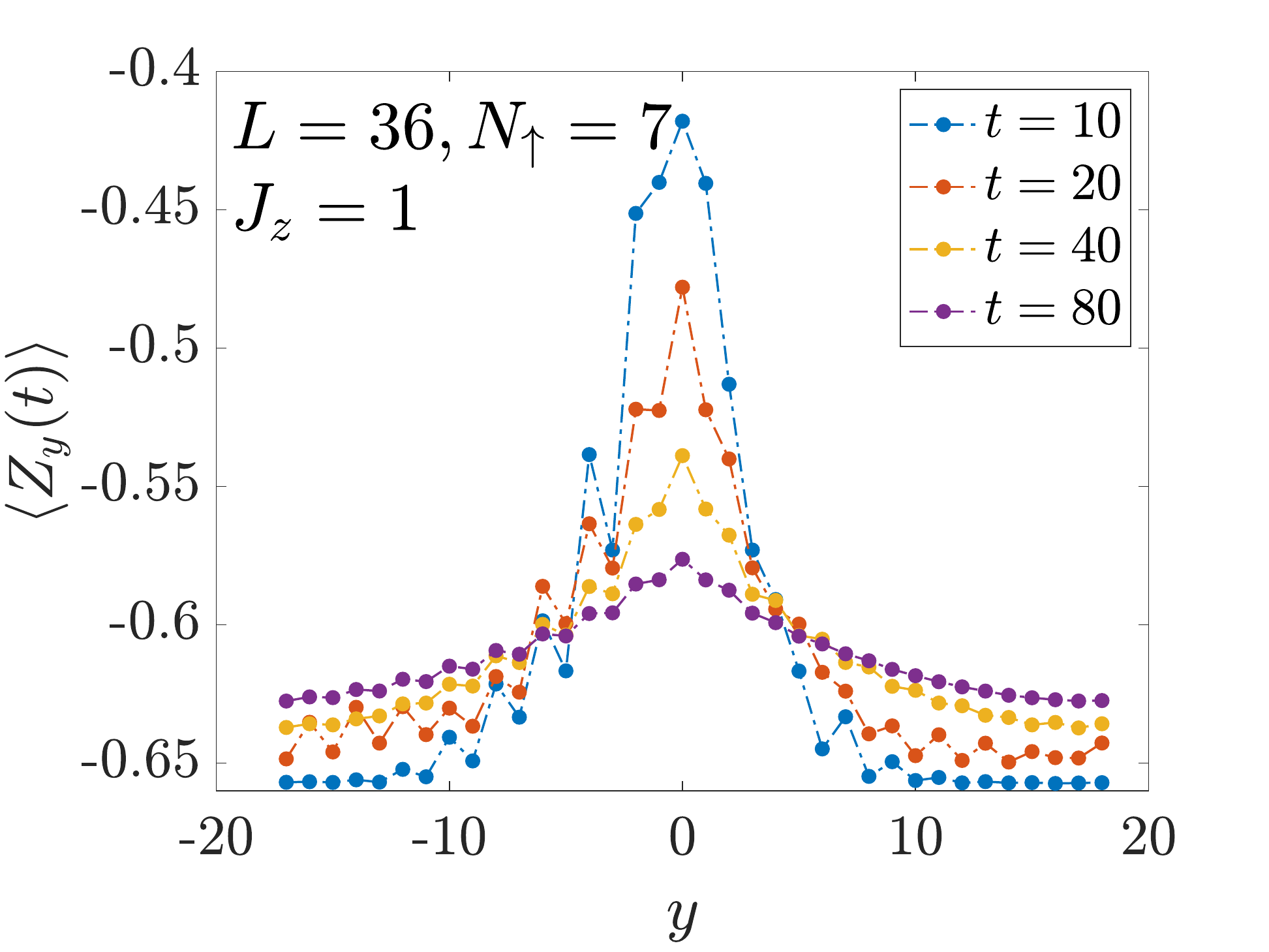}}
\subfigure[]{\label{fig:Z_exp_t} \includegraphics[width=.8\columnwidth]{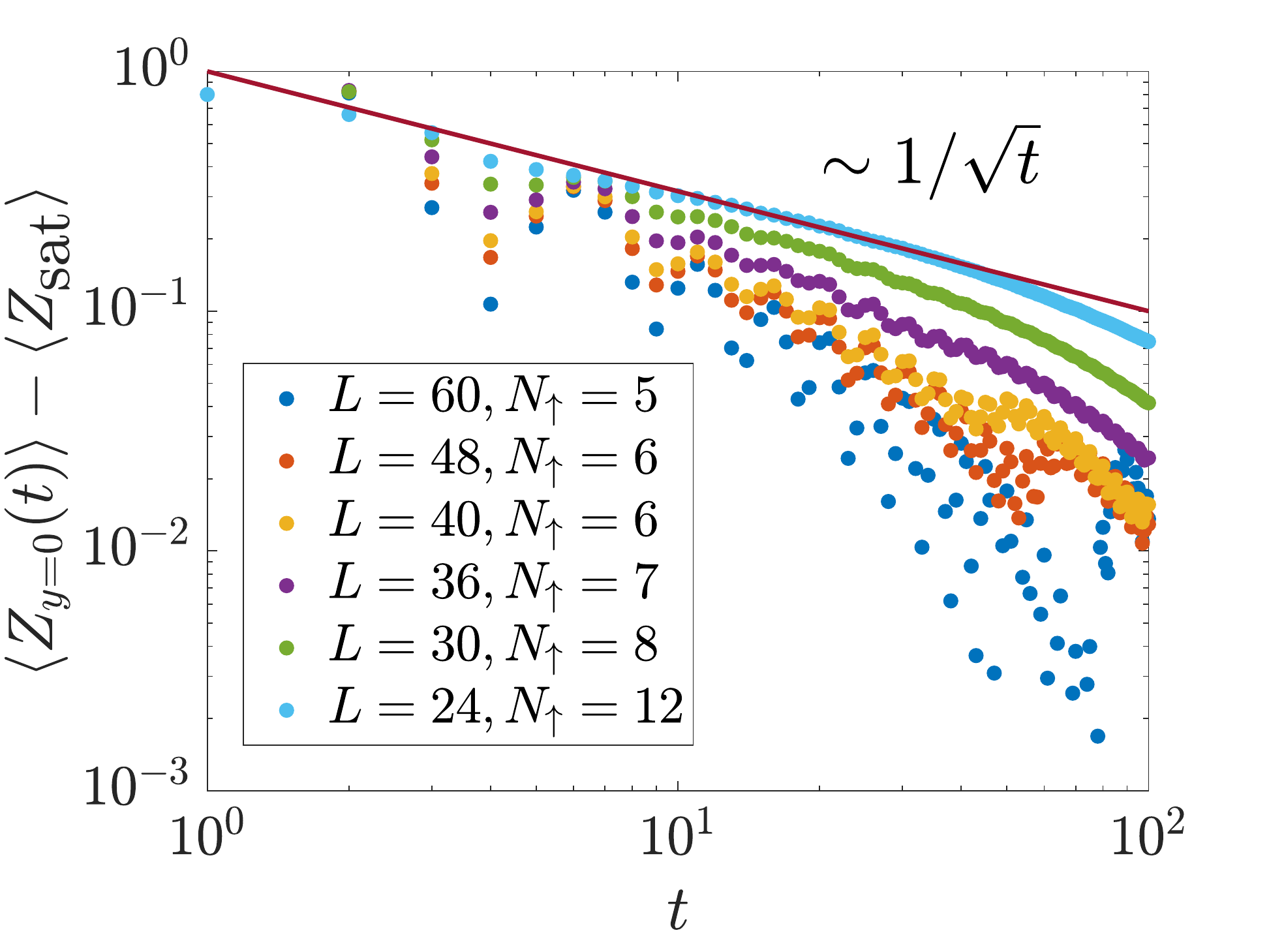}}
\caption{ (a) The $\langle Z_y(t)\rangle$ profile under unitary evolution governed by CFC dynamics with $L=160$ and $N_{\uparrow}=3$. (b) The location of the peak of the right front vs time. Here we only present the results for $t\geq 10$. (c) The $\langle Z_y(t)\rangle$ profile with $L=36$ and $N_{\uparrow}=7$. (d) The dynamics of $\langle Z_y(t)\rangle$ at $y=0$ for various $\alpha$. We observe that it will relax to the saturation value $\langle Z_{\mbox{sat}}\rangle\equiv (2N_{\uparrow}-L)/L$ diffusively when $\alpha$ is large. In all these simulations, the initial state is defined in \eqref{eq:flip_up} with $X^+_x$ applied at the center of the spin chain.} 
\label{fig:Z_exp_dynamics}
\end{figure*}

As shown in  Fig.~\ref{fig:L_160_Jz_1}, when $\alpha$ is small, the spread of $\langle Z_y(t)\rangle$ is ballistic.  The peak of the front moves at a constant velocity which does not depend (strongly) on $\alpha$: see Fig.~\ref{fig:peak_t}.  In contrast, for larger $\alpha$,  as shown in Fig.~\ref{fig:L_36_Jz_1}, the dynamics becomes much slower.   Fig.~\ref{fig:Z_exp_t} further indicates that the relaxation to equilibrium is diffusive, as in the RUC \cite{rakovszky_diffusive_2017,khemani_operator_2017}.  

\begin{figure*}[hbt]
\centering
\subfigure[]{\label{fig:L_160} \includegraphics[width=.8\columnwidth]{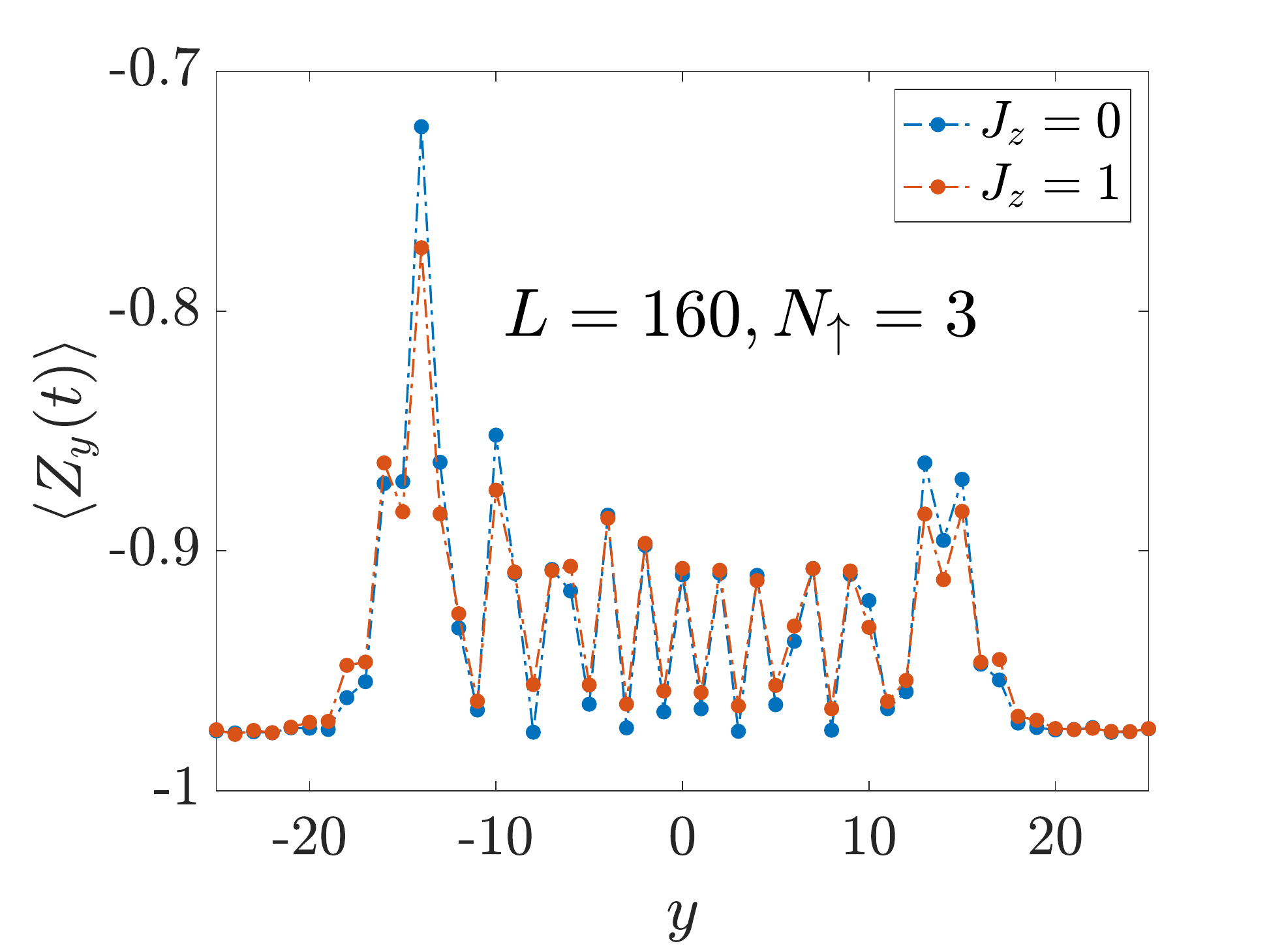}}
\subfigure[]{\label{fig:L_60} \includegraphics[width=.8\columnwidth]{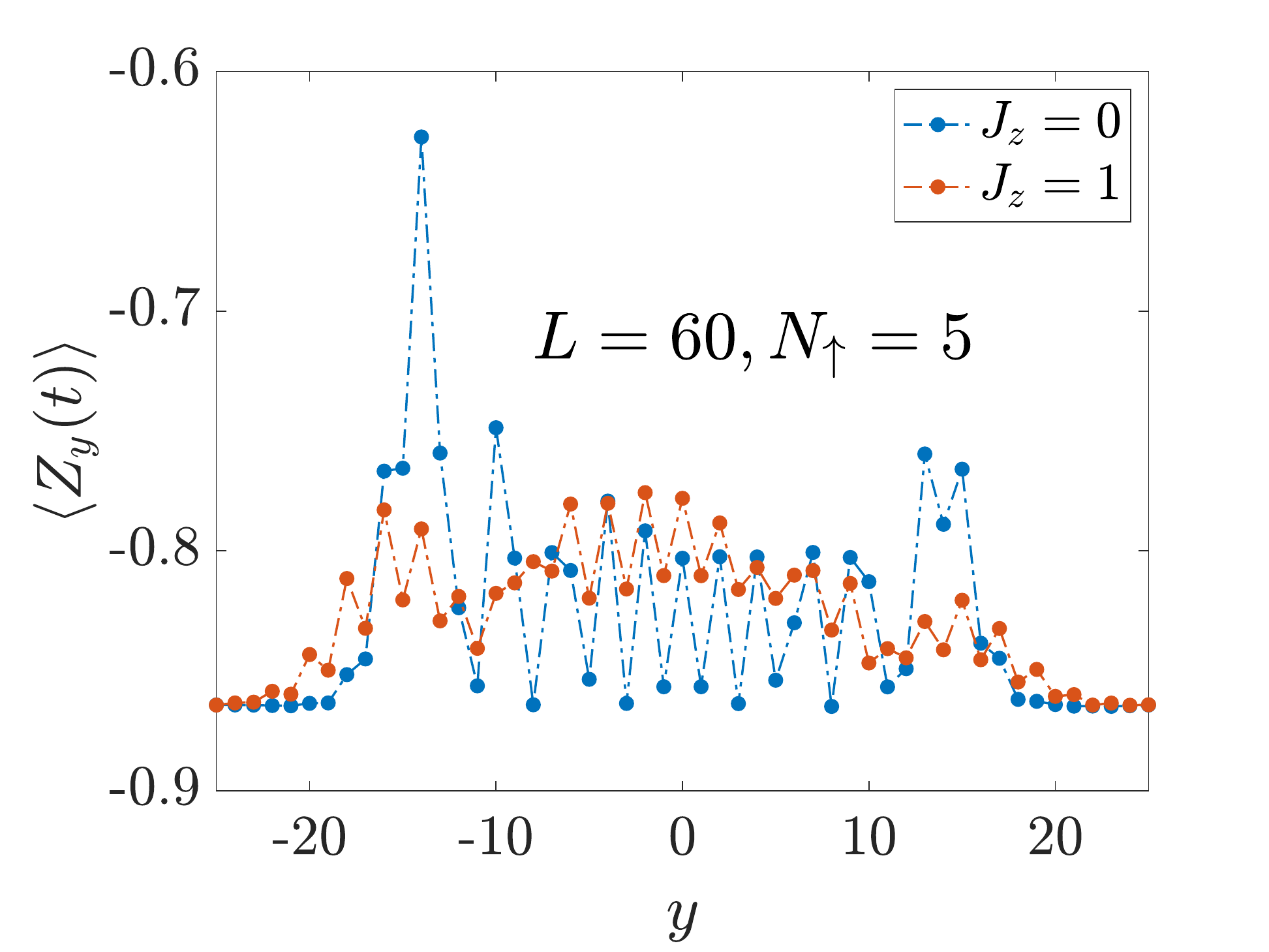}}
\subfigure[]{\label{fig:L_48} \includegraphics[width=.8\columnwidth]{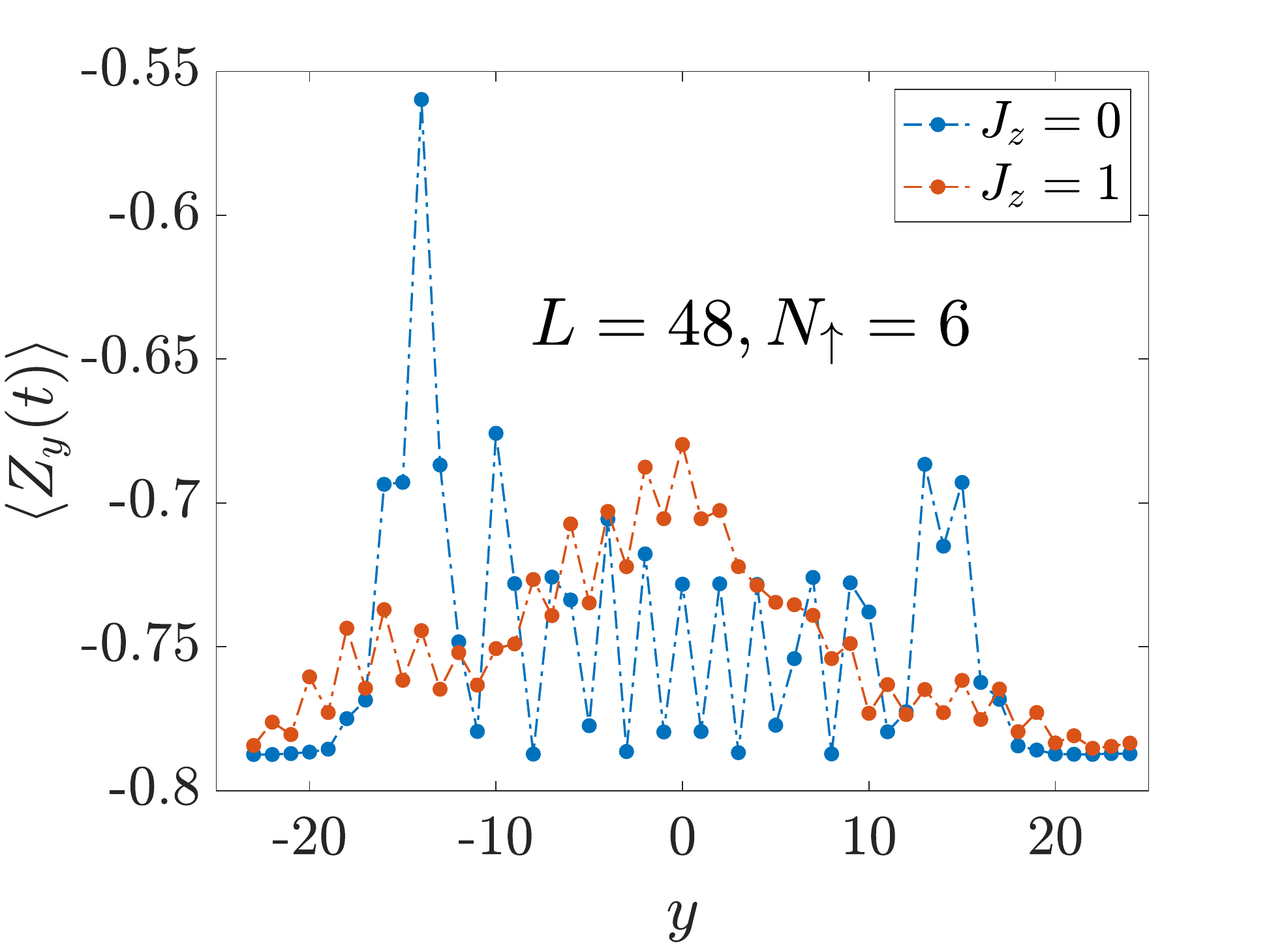}}
\subfigure[]{\label{fig:L_36} \includegraphics[width=.8\columnwidth]{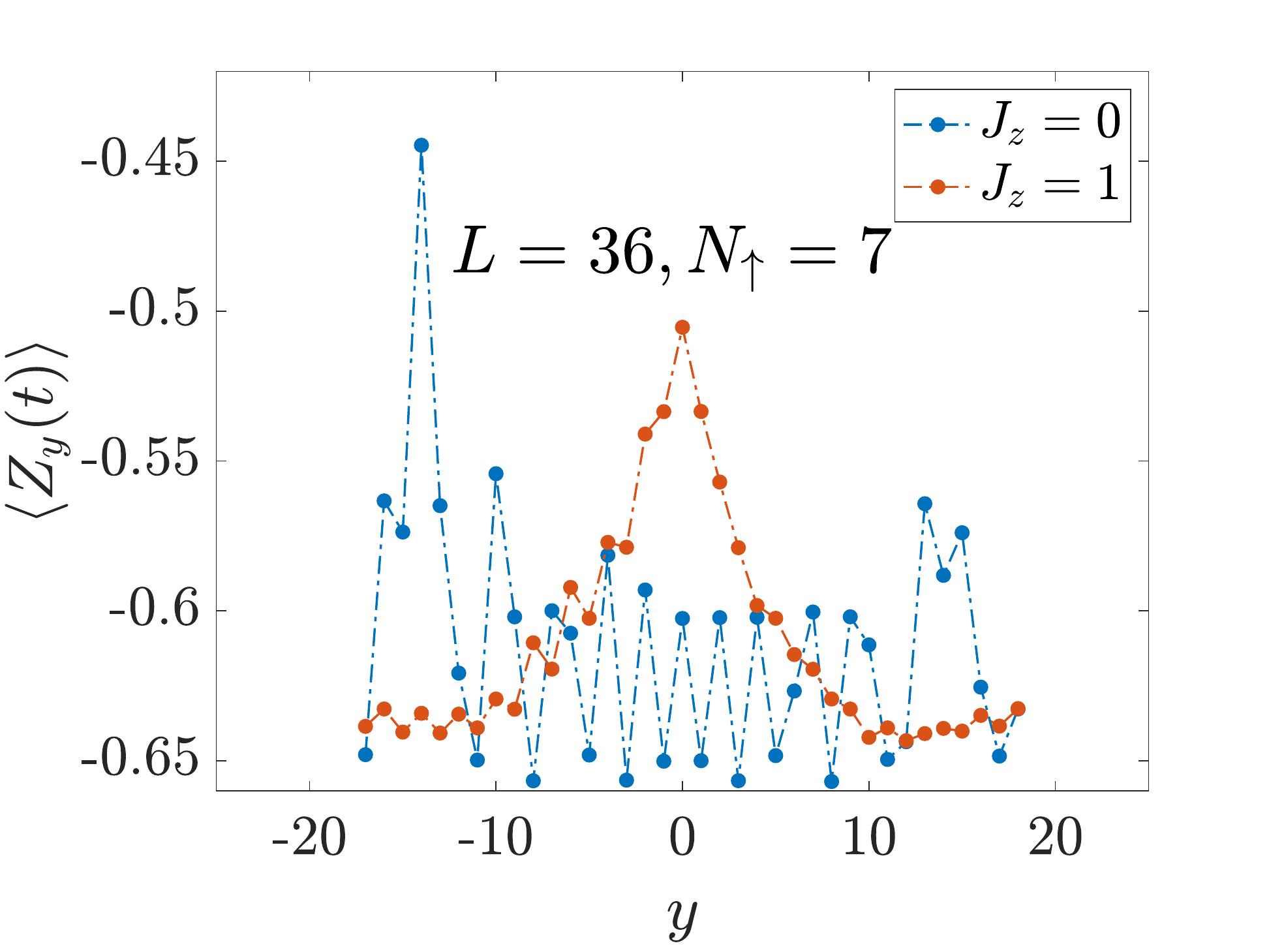}}
\caption{ The numerical results for $\langle Z_y(t)\rangle$ at the same time $t=25$ with different $L$ and $N_{\uparrow}$. All the numerical results are averaged over 10 initial random states defined in \eqref{eq:flip_up}. We notice that the difference between $\langle Z_y(t)\rangle$ of $J_z=0$ and $J_z=1$ increases  as we increase $\alpha\equiv N_{\uparrow}/L$, } 
\label{fig:Z_exp_comparision}
\end{figure*}

Fig.~\ref{fig:Z_exp_comparision} shows $\langle Z_y(t)\rangle$ at a fixed time slice for different values of $J_z$. For $J_z=0$, we observe that $\langle Z_y(t)\rangle$  always spreads ballistically; the speed is independent of $\alpha$. In contrast, for $J_z=1$, as we gradually increase $\alpha$, there is a clear crossover from ballistic to diffusive transport.   While we cannot clearly extract a diffusion constant from the finite size data in Fig.~\ref{fig:Z_exp_t}, the broadening of the diffusive peak at smaller $\alpha$ suggests that the diffusion constant is a decreasing function of $\alpha$. 

\subsection{Cartoon model}

Motivated by the above numerics, we propose a cartoon picture to explain the dynamics of the initially localized up spin and estimate the diffusion constant.  Similarly to before, we model the quantum walk for the right most $X^+$ (labelled $\widehat{X}^+$) in the Pauli string.  The Hilbert space of the quantum walk is spanned by single particle states $|m\rangle$ which label the lattice site of $\widehat{X}^+$.  The unitary time evolution operator is \begin{equation}
    U = \prod_{i=1}^t U_\theta^{n_i} U_2 U_1
    \label{eq:qw_random}
\end{equation}
where $n_i$ is a Bernoulli random variable with \begin{subequations}\begin{align}
    \mathbb{P}[n_i=0]&=1-\alpha, \\
    \mathbb{P}[n_i=1]&=\alpha,
\end{align}
\end{subequations}
the $U_1$ gate is defined separately on even vs. odd sites as:
\begin{subequations}\begin{align}
&U_1|2m-1\rangle=\frac{1}{\sqrt{2}}(|2m-1\rangle+|2m\rangle),\\
&U_1|2m\rangle=\frac{1}{\sqrt{2}}(|2m-1\rangle-|2m\rangle),
\end{align}\end{subequations}
 $U_2$ is 
\begin{subequations}\begin{align}
&U_2|2m\rangle=\frac{1}{\sqrt{2}}(|2m\rangle+|2m+1\rangle),\\
&U_2|2m+1\rangle=\frac{1}{\sqrt{2}}(|2m\rangle-|2m+1\rangle),
\end{align}\end{subequations}
and the phase gate
\begin{align}
U_{\theta}|m\rangle=\mathrm{e}^{\mathrm{i}\theta_m}|m+1\rangle
\label{eq:rp_gate_shift}
\end{align}
where $\theta_m \in [0,2\pi)$ is a uniformly random phase. 

This model can be interpreted as follows:  the $U_1$ and $U_2$ gates generate a single particle coherent quantum walk with ballistic transport.  With probability $\alpha$, the $\widehat{X}^+$ encounters another up spin and the phase gates $U_{Z,1/2}$, which are mimicked by $U_\theta$, dephase the wave function.  Also observe that the dephasing gate $U_\theta$ kicks the $\widehat{X}^+$ to the right, just as in our cartoon of the RUC.

\begin{figure*}[hbt]
\centering
 \subfigure[]{\label{fig:quantum_walk_random_alpha} \includegraphics[width=.8\columnwidth]{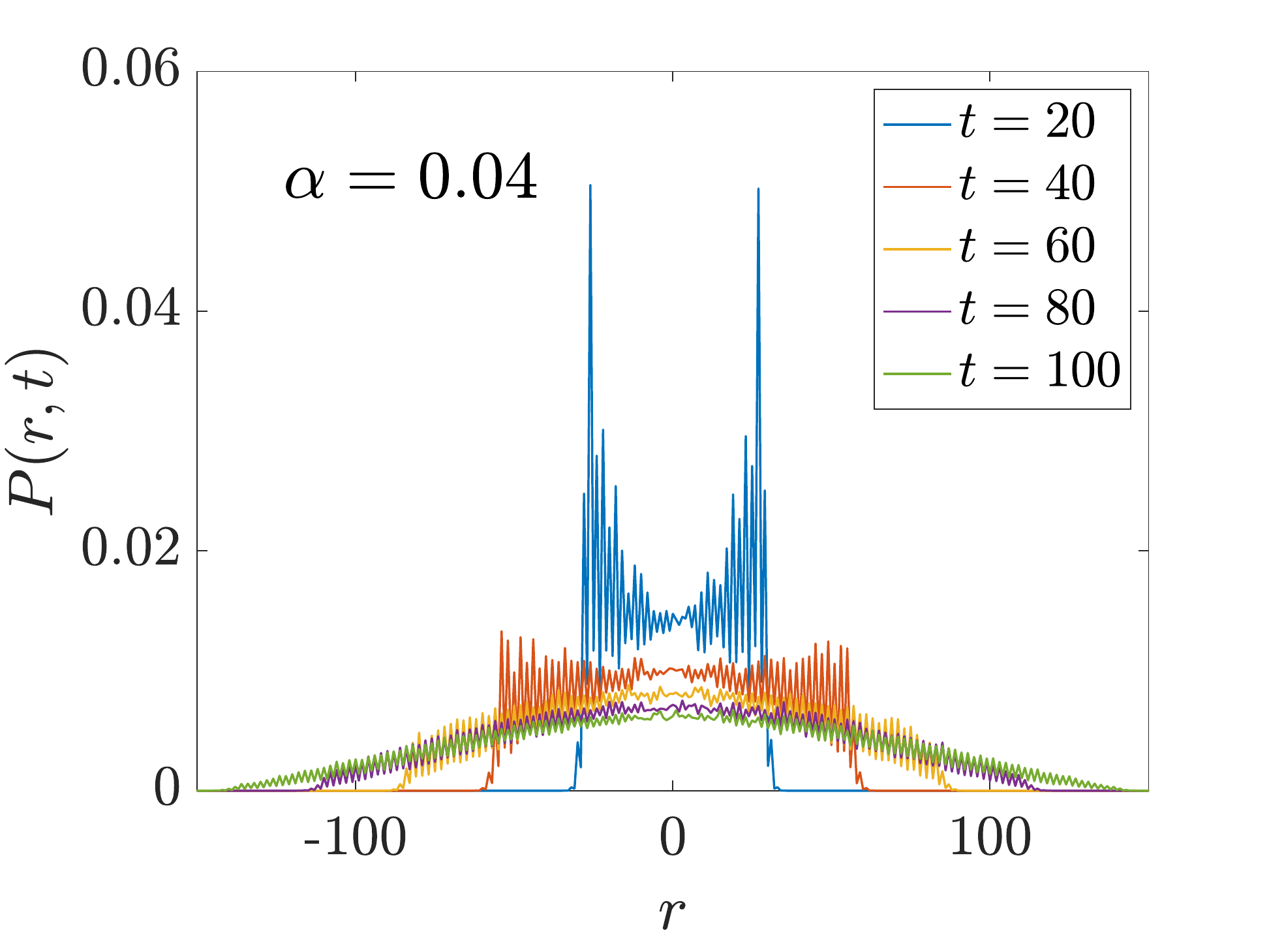}}
 \subfigure[]{\label{fig:symmetric_collapse} \includegraphics[width=.8\columnwidth]{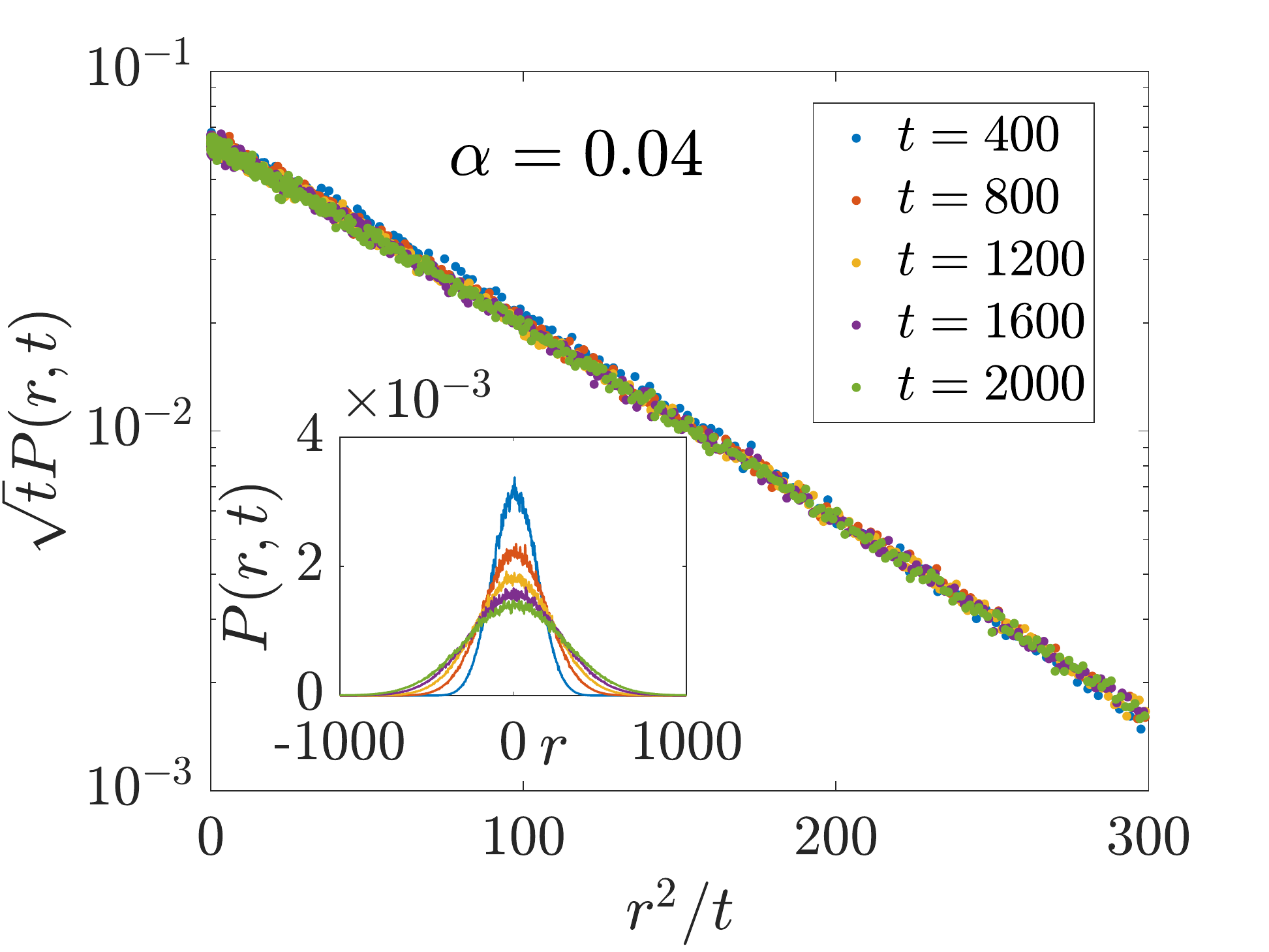}}
  \subfigure[]{\label{fig:Diff_alpha} \includegraphics[width=.8\columnwidth]{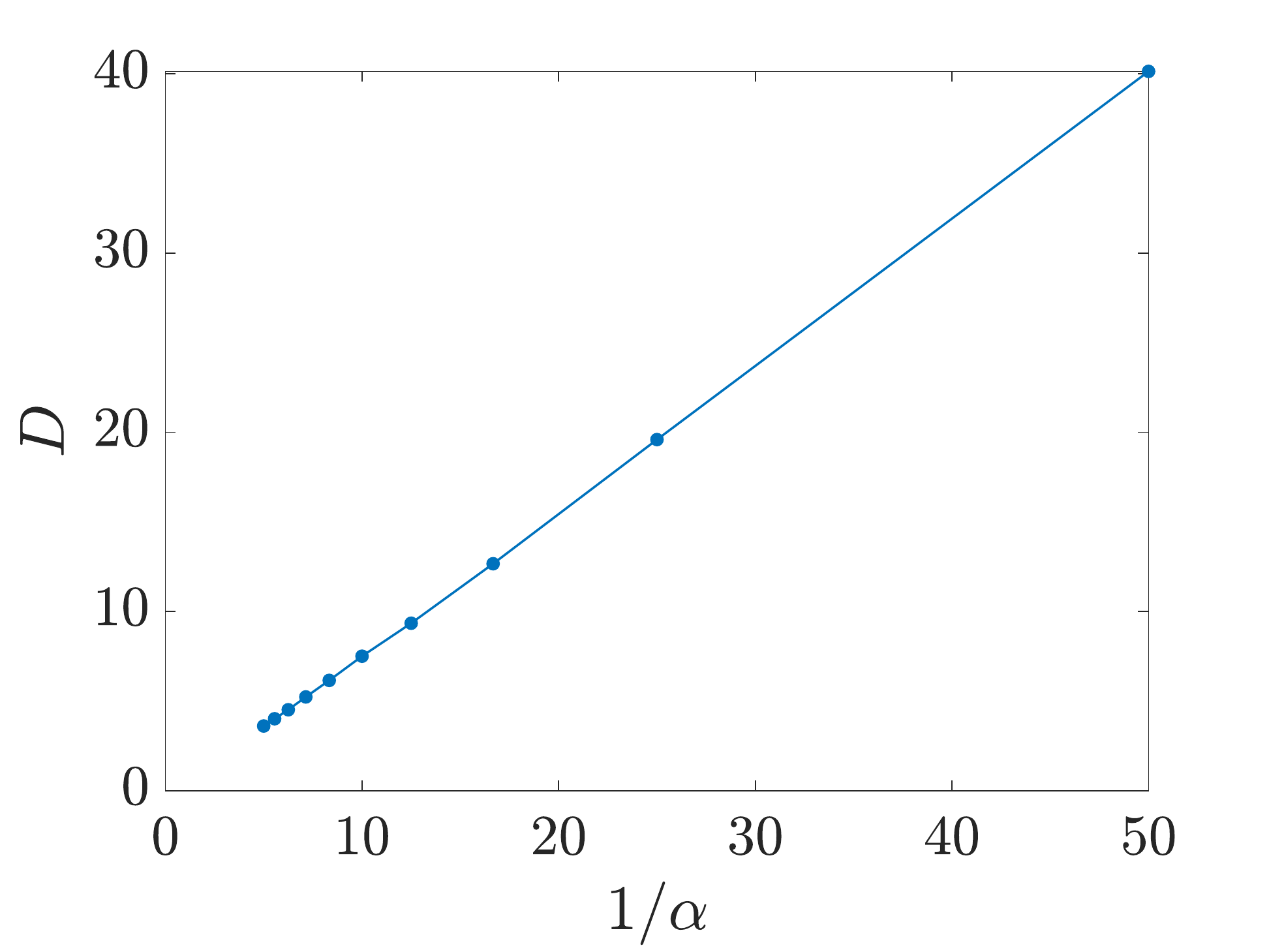}}
  \subfigure[]{\label{fig:quantum_walk_collapse_kick} \includegraphics[width=.8\columnwidth]{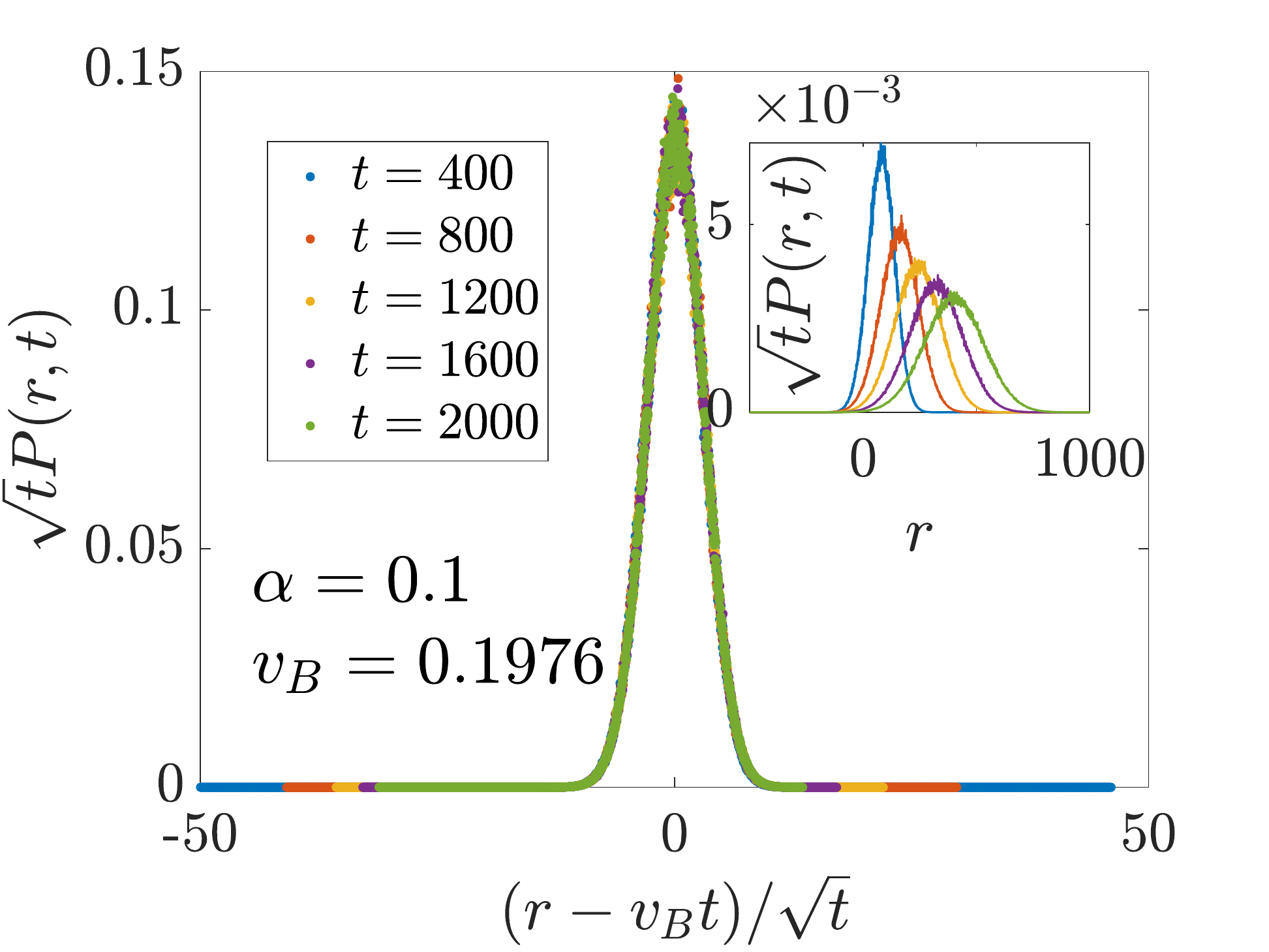}}
\caption{ The numerical results for quantum walk defined in \eqref{eq:qw_random} with random phase gate. The initial state is taken to be $1/\sqrt{2}(|0\rangle+\mathrm{i}|1\rangle)$. (a) We take the random phase gate $U_\theta|m\rangle=\mathrm{e}^{\mathrm{i}\theta_m}|m\rangle$ with probability $\alpha=0.04$ and we observe a crossover from the ballistic to diffusive dynamics. This diffusive dynamics can be further supported by the data collapse in (b). This result indicates that $P(r,t)=\frac{1}{\sqrt{4\pi Dt}}\mathrm{e}^{-\frac{r^2}{4Dt}}$, in which the diffusion constant $D$ can be read out from the slope of the single straight line. (c) We use the same method to extract the diffusion constant $D$ at other $\alpha$ and show that $D\sim 1/\alpha$. (d) We take the random phase gate with a shift defined in \eqref{eq:rp_gate_shift} and perform data collapse for $P(r,t)$. We observe a finite $v_{\mathrm{B}}$ which is linearly proportional to $\alpha$ (not shown in the plot). }
\label{fig:quantum_walk_random}
\end{figure*}

The dynamics of this quantum walk are analyzed in Appendix \ref{app:qw}; here we present a quick argument.  At early times $t\lesssim \alpha^{-1}$, the quantum walk is completely coherent.  After this time scale, $\widehat{X}^+$ experiences a completely dephasing collision.  After the collision, it will take another time $\sim \alpha^{-1}$ to dephase again, etc.  Thus, we conclude that the long time dynamics is a \emph{classical random walk}, but one whose diffusion constant is parametrically large:  $D = \Delta x^2 / \Delta t$, where $\Delta x \sim \alpha^{-1}$ and $\mathrm{\Delta }t \sim \alpha^{-1}$ are the length and time step for the classical random walk.  In other words, \begin{equation}
    D\sim \alpha^{-1}.
\end{equation} 
Moreover, the butterfly velocity $v_{\mathrm{B}}$ is controlled only by the biased rightwards motion of the dephasing collisions: \begin{equation}
    v_{\mathrm{B}}\sim \frac{1}{\Delta t} \sim \alpha.
\end{equation} 
This argument is easily observed numerically in simulations of this cartoon model: see Fig.~\ref{fig:quantum_walk_random}.


Interestingly, $v_{\mathrm{B}}$ is quite similar in the CFC and the RUC, essentially because only the collisions of two up spins contribute to operator growth.  The diffusion constant, which is physically (more) measurable than operator growth, parametrically differs from the CFC to the RUC, as a consequence of coherent quantum dynamics on short time and length scales.  We conclude that the RUC only quantitatively models operator dynamics in the ``infinite temperature" sector $\alpha\approx 1/2$.  These conclusions do not depend on details of the model we studied: it is easy to see that the coherent quantum dynamics that differ from the noisy RUC dynamics at short times are universal properties of any time-independent Floquet system or Hamiltonian system.

\section{Outlook}
We have compared the physical predictions for operator growth between random circuit models and non-random chaotic Floquet models with conservation laws in one dimension.  We find that $v_{\mathrm{B}} \sim \alpha$ is consistent between both models, but the diffusion constant $D\sim \alpha^0$ vs. $D\sim \alpha^{-1}$ is parametrically different.   This presents a very simple, and physically realistic, example of chaotic quantum dynamics which is only partially captured by random circuits.  (See also the more pathological ``star graph" model of \cite{ Lucas2019_star}.)    It is important to understand whether the predictions of exotic dynamics in other classes of random circuits are properties of chaotic Floquet or Hamiltonian evolution without randomness, or are peculiar features of  time-dependent randomness.

We expect that in most physical models, the dominant slow down of the butterfly velocity at high polarization, or low temperature, is arising from quantum dephasing between growing operators.    So it is unclear whether current techniques based on  the Lieb-Robinson theorem \cite{Lieb1972}  or beyond \cite{Chen:2019hou,Chen:2019klo}  provide non-trivial constraints on constrained quantum dynamics in generic models \cite{Han:2018bqy}.  

Lastly, it has been argued on rather general grounds that \cite{Hartman:2017hhp, Lucas:2017ibu}\begin{equation}
    D \lesssim v_{\mathrm{B}}^2 \tau_{\mathrm{hydro}}  \label{eq:hartnollineq}
\end{equation} is a requirement of causality;  here $\tau_{\mathrm{hydro}}$ corresponds to the time scale beyond which hydrodynamics is a sensible effective theory.   This inequality comes from demanding that the diffusive front $x\sim \sqrt{Dt}$ lies within the operator light cone.  In the RUC, (\ref{eq:hartnollineq}) makes sense as written: $D\sim \alpha^0$, $v_{\mathrm{B}}\sim \alpha$ and $\tau_{\mathrm{hydro}} \sim \alpha^{-2}$ is set by the decay rate of a diffusion mode on the length scale $\alpha^{-1}$ where a single excitation (up spin) is present.   However, if we apply (\ref{eq:hartnollineq}) to the CFC, it appears that $\tau_{\mathrm{hydro}} \gtrsim \alpha^{-3}$ -- namely, hydrodynamics breaks down at anomalously late times \cite{Lucas:2017ibu}.  In this model, however, this reasoning is incorrect.   At early times $t< \alpha^{-1}$, in the CFC $v_{\mathrm{B}}\sim \alpha^0$ due to the single particle quantum walk.  Hence, (\ref{eq:hartnollineq}) should be interpreted as $\alpha^{-1} \lesssim (\alpha^0)^2 \alpha^{-1}$. In other words, the late time butterfly velocity ends up irrelevant for diffusion, which is controlled entirely by single particle dynamics.

\acknowledgements
We acknowledge useful discussion with Curt von Keyserlingk and Tibor Rakovszky. XC and RMN acknowledge support from the DARPA DRINQS program.

\appendix 
\section{Single particle diffusion constant}\label{app:1partD}
Take an initial state with only one spin pointing up, i.e., $|\psi_0\rangle=|\downarrow\ldots\downarrow\downarrow\uparrow\downarrow\downarrow\ldots\downarrow\rangle$, under the random circuit described in Fig.~\ref{fig:random_circuit} (a), the up spin is performing a classical random walk. In this section, we compute the diffusion constant for this random walker.

We first consider a two-qubit state $|\uparrow\downarrow\rangle$, under $U_2$ gate, it becomes superposition of $|\uparrow\downarrow\rangle$ and $|\downarrow\uparrow\rangle$ with the same probability $1/2$ on average, i.e.,
\begin{align}
& \mathbb{E}[|\langle \uparrow\downarrow|U_2|\uparrow\downarrow\rangle|^2]=\frac{1}{2}\nonumber\\
&\mathbb{E}[|\langle \uparrow\downarrow|U_2|\downarrow\uparrow\rangle|^2]=\frac{1}{2},
\end{align}
where $\mathbb{E}[\cdot]$ denotes average over random $U_2$ gate. Notice that 
\begin{align}
\mathbb{E}[\langle \uparrow\downarrow|U_2|\uparrow\downarrow\rangle]=\mathbb{E}[\langle \uparrow\downarrow|U_2|\downarrow\uparrow\rangle]=0.
\end{align}
Therefore this quantum dynamics can be mapped to a classical random walk. 

For a state $|\psi_0\rangle=|\downarrow\ldots\downarrow\downarrow\uparrow\downarrow\downarrow\ldots\downarrow\rangle$, if the up spin is at odd site $n$, after one time step (the dashed block in Fig.~\ref{fig:random_circuit} (a)), (1) this up spin can move to $n+2$, $n\pm 1$ and $n$ with the same probability $1/4$. Similarly, for a up spin initially at even site $n$, after one time step, (1) this up spin can move to $n-2$, $n\pm 1$ and $n$ with the same probability $1/4$.

We define the probability that the up spin is at site $n$ as $P(n,t)$. The first moment (the mean displacement) and the second moment (variance) at each time is
\begin{align}
& M_1(t)=\sum_n n P(n,t)\\
& M_2(t)=\sum_n n^2 P(n,t).
\end{align}
After evolving for one time step, the first moment becomes
\begin{align}
&M_1(t+1)=\sum_n n P(n,t+1)\nonumber\\
=&\sum_n (2n+1+\frac{1}{2}) P_o(2n+1,t)+(2n-\frac{1}{2}) P_e(2n,t)\nonumber\\
=&\sum_n nP(n,t)+\frac{1}{2}\sum_n(P_o(2n+1,t)-P_e(2n,t))=M_1(t)
\end{align}
where $P_o(2n+1,t)$ and $P_e(2n,t)$ denote the probability at odd and even sites respectively. Notice that $P_e(2n,t)=P_o(2n+1,t)$ at any arbitrary time $t>0$ (independent of the initial state) and we have the first moment is invariant under time evolution. Furthermore, we can show
\begin{align}
&\sum_n (2n+1)P_o(2n+1,t+1)=\frac{M_1(t)}{2}+\frac{1}{4}\nonumber\\
&\sum_n (2n)P_e(2n,t+1)=\frac{M_1(t)}{2}-\frac{1}{4}.
\end{align}
The second moment satisfies
\begin{align}
& M_2(t+1)=\sum_n n^2 P(n,t+1)\nonumber\\
=& \frac{1}{4}\sum_n P_o(2n+1,t) \left[(2n+1)^2 + (2n+1+1)^2 \right.\nonumber\\
&+\left. (2n+1-1)^2 + (2n+1+2)^2 \right]\nonumber\\
&+ \frac{1}{4}\sum_n P_e(2n,t) \left[(2n)^2 + (2n+1)^2 \right.\nonumber\\
&+\left. (2n-1)^2 + (2n-2)^2 \right]\nonumber\\
=&M_2(t)+\frac{3}{2}+\sum_n\left[ (2n+1)P_o(2n+1,t)-2n P_e(2n,t)\right]\nonumber\\
=&M_2(t)+2.
\end{align}
Therefore we have $M_2(t+1)-M_2(t)=2$ and the diffusion constant for the random walker is $D=1$ (since $M_2(t) = 2Dt$ by convention).


\section{Operator growth in the Haar RUC} \label{app:rucrules}
A Pauli string operator $O$ acting on two adjacent qubits transitions according to the following rules under the Haar RUC. 
\begin{itemize}  
\item The following operators are invariant under the RUC: \begin{enumerate}
    \item $P^\downarrow P^\downarrow$, $ P^\uparrow P^\uparrow$, $P^\downarrow P^\uparrow + P^\uparrow P^\downarrow$
    \item $X^+ X^+$, $X^- X^-$
\end{enumerate}  
\item In each set below, operators (which we have not normalized) mix amongst themselves with equal probability:
\begin{enumerate}
    \item $P^\downarrow P^\uparrow -P^\uparrow P^\downarrow, X^+X^-, X^- X^+$
    \item $P^\uparrow X^-, X^- P^\uparrow$
    \item $P^\downarrow X^+, X^+ P^\downarrow$
    \item $P^\downarrow X^-, X^- P^\downarrow$
    \item $P^\uparrow X^+, X^+ P^\uparrow$
    \end{enumerate}
\end{itemize}

\section{Operator growth in the quantum automaton RUC} \label{app:automaton}
The Pauli string operator $O$ defined on three qubits can be classified into two classes according to the adjoint action of gate $U_3$:
\begin{itemize}
\item These operators are invariant under $U_3^\dag O U_3$: 
\begin{enumerate}
\item $P^{\uparrow,\downarrow}P^{\uparrow}P^{\uparrow,\downarrow}$, $P^\uparrow P^\downarrow P^\uparrow$, $P^\downarrow P^\downarrow P^\downarrow$

\item $P^\uparrow X^\pm P^\uparrow$, $P^\downarrow X^\pm P^\downarrow$

\item $X^\pm P^{\uparrow}P^{\uparrow,\downarrow}$, $P^{\uparrow,\downarrow} P^{\uparrow}X^{\pm}$

\item $X^\pm P^\uparrow X^\pm$, $X^+ P^\downarrow X^+$, $X^- P^\downarrow X^-$

\item $P^\uparrow X^+ X^-$, $P^\downarrow X^+ X^+$, $ X^+ X^+ P^\downarrow$, $ X^- X^+ P^\uparrow$
$P^\uparrow X^- X^+$, $P^\downarrow X^- X^-$, $ X^- X^- P^\downarrow$, $ X^+ X^- P^\uparrow$

\item $X^+X^+X^+$, $X^-X^-X^-$, $X^-X^+X^-$, $X^+X^-X^+$
\end{enumerate}
\item These operators will transform in the following way:
\begin{enumerate}
\item $P^\uparrow P^\downarrow P^\downarrow\longleftrightarrow P^\downarrow P^\downarrow P^\uparrow$, 

\item $P^\uparrow X^+ P^\downarrow \longleftrightarrow X^+ X^+ X^-$, \\
$\quad$ $P^\downarrow X^+ P^\uparrow \longleftrightarrow X^- X^+ X^+$, \\
$P^\uparrow X^- P^\downarrow \longleftrightarrow X^- X^- X^+$, \\
$P^\downarrow X^- P^\uparrow \longleftrightarrow X^+ X^- X^-$

\item $X^\pm P^{\downarrow} P^{\uparrow,\downarrow} \longleftrightarrow P^{\uparrow,\downarrow} P^\downarrow X^\pm$,

\item $X^+P^\downarrow X^- \longleftrightarrow X^-P^\downarrow X^+$, 

\item $P^\uparrow X^+ X^+\longleftrightarrow X^+X^+P^\uparrow$, \\
$P^\uparrow X^- X^-\longleftrightarrow X^-X^-P^\uparrow$\\
$P^\downarrow X^+ X^-\longleftrightarrow X^-X^+P^\downarrow$, \\
$P^\downarrow X^- X^+\longleftrightarrow X^+X^-P^\downarrow$\\
\end{enumerate}
\end{itemize}

\section{Operator dynamics for quantum walk on a line} \label{app:qw}
\subsection{Discrete time quantum walk}
In this section, we study a class of Floquet circuit with $\mathrm{U}(1)$ symmetry. We show that if the initial state only has one up spin with the rest of spins pointing down, the dynamics for the up spin corresponds to a quantum walk problem \cite{Kempe2003}. The Floquet operator is defined as
\begin{align}
U_F=\left(\prod_m U_{2m,2m+1}^{(2)}\right)\left(\prod_m U_{2m-1,2m}^{(1)}\right)
\end{align}
with $U^{(1)}_{2m-1,2m} $ and $U^{(2)}_{2m,2m+1}$ defined on two neighboring qubits. They preserve $\mathrm{U}(1)$ symmetry and take the following form,
\begin{align}
U^{(1)/(2)}\equiv\begin{pmatrix}
1 & & \\
& U_{\uparrow\downarrow}^{(1)/(2)} & \\
& & 1
\end{pmatrix}
\end{align}
where $U_{\uparrow\downarrow}^{(1)/(2)}$ is a $2\times 2$ unitary matrix defined on the subspace composed by $|\uparrow\downarrow\rangle$ and $|\downarrow\uparrow\rangle$. We define the probability for the up spin at odd site as $a_m(t)$ and even site as $b_m(t)$ with the constraint $\sum_m |a_m(t)|^2+|b_m(t)|^2=1$. Under Floquet operator, we have
\begin{align}
&\begin{pmatrix}
a_m(t+\frac{1}{2})\\
b_m(t+\frac{1}{2})
\end{pmatrix}=U_{\uparrow\downarrow}^{(1)}\begin{pmatrix}
a_m(t)\\
b_m(t)
\end{pmatrix}\\
&\begin{pmatrix}
b_m(t+1)\\
a_{m+1}(t+1)
\end{pmatrix}=U_{\uparrow\downarrow}^{(2)}\begin{pmatrix}
b_m(t+\frac{1}{2})\\
a_{m+1}(t+\frac{1}{2})
\end{pmatrix}
\end{align}
The above equation characterizes the discrete quantum walk process and can be convenient rewritten  in the momentum space. We define the Fourier transformation, 
\begin{align}
a_k(t)=\frac{1}{\sqrt{L}}\sum_m e^{-i m k}a_m(t),\ b_k(t)=\frac{1}{\sqrt{L}}\sum_m e^{i m k}b_m(t),
\end{align}
and we have
\begin{align}
\begin{pmatrix}
a_k(t+1)\\
b_k(t+1)
\end{pmatrix}=M_k\begin{pmatrix}
a_k(t)\\
b_k(t)
\end{pmatrix},
\end{align}
where $M_k$ is $2\times 2$ unitary matrix. The distribution of $a_m(t)$ and $b_m(t)$ can be studied by diagonalizing $M_k$ matrix and performing the inverse Fourier transformation.

We consider a simple example with
\begin{align}
U^{(1)}=\frac{1}{\sqrt{2}}\begin{pmatrix}
1 & 1\\
1 & -1
\end{pmatrix},\  U^{(2)}=\frac{1}{\sqrt{2}}\begin{pmatrix}
0 & -1\\
1 & 0
\end{pmatrix},
\end{align}
so that 
\begin{align}
M_k=\frac{1}{\sqrt{2}}\begin{pmatrix}
-e^{-ik} & e^{ik} \\
e^{-ik} & e^{ik}
\end{pmatrix}.
\end{align}
This actually is the famous Hadamard quantum walk and has been analytically studied in Ref.\ \onlinecite{ambainis2001}.  Below we briefly review their main results. If we start with an initial state with the amplitude $a_0=1$ and rest of them equal to zero, as time evolves, the position probability of the up spin will spread out rapidly. At time $t$, $|a_m(t)|^2$ and $|b_m(t)|^2$ will  become roughly uniform in the interval $[-t/\sqrt{2},t/\sqrt{2}]$, with the variance $\sigma^2\sim t^2$. Outside the interval, it dies out quickly. At the fronts at $\pm t/\sqrt{2}$, there are two peaks with width $t^{-1/3}$. This ballistic spreading is in contrast with the classical random walk where we have $\sigma^2\sim t$. 
\subsection{Quantum walk in random medium}
The ballistic spreading in quantum walk is due to quantum coherence and is unstable if we introduce decoherent events. In this section, we introduce unitary random phase gate into quantum walk and study the transition to the classical random walk in the long time limit.
The method we are using here is following Ref.\ \onlinecite{Romanelli2005}, where they were considering the dephasing effect due to the non-unitary projective measurement.
Consider an initial product state $|\psi_0\rangle=|\downarrow\downarrow\ldots\downarrow\uparrow\downarrow\ldots\downarrow\downarrow\rangle$, under unitary time evolution, the up spin performs one dimensional ballistic quantum walk and the wave function can be written as
\begin{align}
|\psi(t)\rangle=\sum_n c_n(t)|n\rangle
\label{eq:wf},
\end{align}
where $|n\rangle$ denotes the state with spin at site $n$ pointing up. 
For this quantum walk model, after involving it for time $t$, we apply a random phase gate 
\begin{align}
U_{\theta}|n\rangle=e^{i\theta_n}|n\rangle,
\label{eq:rp_gate}
\end{align}
where $\theta_n\in [0,2\pi]$ is a random phase. This random phase removes the quantum coherence in the quantum wave function \eqref{eq:wf}. The probability that the spin is pointing up at site $n$ is $P_n(t)=|c_n(t)|^2$. Notice that $P_n$ satisfies the constraint $\sum_n P_n=1$. Below we are going to study how $P_n(t)$ spreads out in time. We average over the random phase gate, as in Appendix \ref{app:1partD}.

The first and second moments are defined as
\begin{align}
&M_1(t)=\sum_n nP_n(t)\\
& M_2(t)=\sum_n n^2P_n(t).
\end{align}
Furthermore, the variance is defined as
\begin{align}
\sigma^2(t)\equiv M_2(t)-M_1^2(t).
\end{align}

After applying this random phase gate, the quantum interference between different $|n\rangle$ is lost. We continue to evolve the wave function with quantum walk gate for time $T$. The probability distribution at time $t+T$ is 
\begin{align}
P_n(t+T)=\sum_{j=n-2T}^{n+2T} P_{n-j}(T)P_j(t).
\end{align}

At this time, the first moment has
\begin{align}
&M_1(t+T)=\sum_n nP_n(t+T)=\sum_n n \sum_{j=n-2T}^{n+2T} P_{n-j}(T)P_j(t)\nonumber\\
&=\sum_j \sum_{l=-2T}^{2T}(j+l)P_j(t)P_l(T)=\sum_j j P_j(t) +\sum_{l=-2T}^{2T} l P_l(T)\nonumber\\
&=M_1(t)+ M_1(T).
\end{align}

The second moment has
\begin{align}
&M_2(t+T)=\sum_n n^2P_n(t+T)=\sum _j \sum_{l=-2T}^{2T}(j+l)^2P_j(t)P_l(T)\nonumber\\
&=M_2(t)+M_2(T)+2M_1(t)M_1(T).
\end{align}

Therefore we have the variance $\sigma^2(t+T)$ satisfies
\begin{align}
\sigma^2(t+T)\equiv M_2(t+T)-M_1^2(t+T)=\sigma^2(t)+\sigma^2(T).
\end{align}

Using the above result, we are ready to consider a quantum circuit model, in which we apply $U_{\theta}$ gate at random times with time intervals $T_1, T_2, T_3\ldots$. In this case, the wave function evolves as
\begin{align}
|\psi\rangle=\ldots U(T_3)U_\theta U(T_2)U_\theta U(T_1)|\psi_0\rangle.
\end{align}
The total variance $\sigma^2(\sum_iT_i)=\sum_i T_i^2$, and the diffusion constant is
\begin{align}
D=\frac{1}{2}\frac{{\sigma^2(\sum_i T_i)}}{\overline{\sum_iT_i}}\sim \overline{T}\sim \frac{1}{\alpha},
\end{align}
where $\overline{T}$ is the mean time interval. 
We can extend the above result to the random phase gate with a shift, i.e.,
\begin{align}
U_{\theta}|n\rangle=e^{i\theta_n}|n+1\rangle.
\label{eq:rp_gate_kick}
\end{align}
In this case, we have the biased random walk with the same diffusion constant and butterfly velocity $v_{\mathrm{B}}\sim \alpha$.

\bibliography{quantum_walk}
\end{document}